\documentclass{amsart}
\usepackage[utf8]{inputenc}
\usepackage{amssymb, amsmath, mathtools, amsthm,subcaption}
\usepackage[all]{xy}
\usepackage{fullpage}
\usepackage{adjustbox}
\usepackage{color}

\newcommand{\bC}{{\mathbb C}}

\newcommand{\bZ}{{\mathbb Z}}

\newcommand{\cH}{{\mathcal H}}

\newcommand{\Ir}{\mathbb{Z}} 
\newcommand{\Rl}{\mathbb{R}} 
\newcommand{\Cx}{\mathbb{C}}
 
\newcommand{\caG}{{\mathcal G}}
\newcommand{\caH}{{\mathcal H}}

\newcommand{\bbZ}{{\mathbb Z}}

\newcommand{\braket}[2]{\left\langle #1 , #2\right\rangle}

\newcommand{\ketbra}[1]{\vert #1\rangle\langle #1\vert}

\newcommand{\ket}[1]{\vert #1 \rangle}

\newcommand{\spec}{\mathrm{spec}}

\newtheorem{theorem}{Theorem}[section]
\newtheorem{lemma}[theorem]{Lemma}

\newtheorem{corollary}[theorem]{Corollary} 
\newtheorem{remark}[theorem]{Remark}
\newcommand{\be}{\begin{equation}}
\newcommand{\ee}{\end{equation}}
\newcommand{\bea}{\begin{eqnarray}}
\newcommand{\eea}{\end{eqnarray}}
\newcommand{\beann}{\begin{eqnarray*}}
\newcommand{\eeann}{\end{eqnarray*}}
\newcommand{\eq}[1]{(\ref{#1})}
\definecolor{Green}{rgb}{0,.8,.4}
\definecolor{Plum}{rgb}{.5,0,1}
\definecolor{Orange}{rgb}{1,.6,0}

\begin{document}
\title{Spectral Gap and Edge Excitations\\ of $d$-dimensional PVBS models on half-spaces}
\date{\today}

\author{Michael Bishop}
\address{Department of Mathematics, University of California, Davis,
Davis, CA 95616}
\email{mbishop@math.ucdavis.edu}
\author{Bruno Nachtergaele}
\address{Department of Mathematics, University of California, Davis,
Davis, CA 95616}
\email{bxn@math.ucdavis.edu}
\author{Amanda Young}
\address{Department of Mathematics, University of California, Davis,
Davis, CA 95616}
\email{amyoung@math.ucdavis.edu}

\begin{abstract}
We analyze a class of quantum spin models defined on half-spaces in the 
$d$-dimensional hypercubic lattice bounded by a hyperplane with inward
unit normal vector $m\in\Rl^d$. The family of models was previously introduced 
as the single species Product Vacua with Boundary States (PVBS) model, which
is a spin-$1/2$ model with a XXZ-type nearest neighbor interactions depending 
on parameters $\lambda_j\in (0,\infty)$, one for each coordinate direction.
For any given values of the parameters, we prove an upper bound for the
spectral gap above the unique ground state of these models, which vanishes
for exactly one direction of the normal vector $m$. For all other choices of $m$
we derive a positive lower bound of the spectral gap, except for the case $\lambda_1
=\cdots =\lambda_d=1$, which is known to have gapless excitations in the bulk.
\end{abstract}

\maketitle

\section{Introduction} \label{sec:Introduction}

One of the essential properties to understand the low-temperature behavior of 
a quantum lattice model is the presence or absence of gapless excitations above
the ground state or, equivalently, whether or not there is a non-vanishing spectral
gap above the ground state. Even when the ground state is known, answering
this question is, in general, quite non-trivial, especially in higher dimensions.
We are interested in developing techniques to prove lower bounds for the spectral
that work in two and more dimensions by studying specific models.

The issue of the existence of a spectral gap may be further complicated by the 
presence of gapless edge states, which can occur for certain geometries 
while the excitations in the bulk remain gapped.
Edge states play a central role in characterizing quantum many-body states.
The occurrence of low-energy, often gapless, states supported near the boundary
of an extended many-body system have been connected with entanglement
properties and topological order \cite{qi:2012} and with phenomena such as
the quantum Hall effect and the spin Hall effect. They may reflect the correlation 
structure of the state in the bulk and have a direct bearing on the classifcation of
gapped ground state phases and the phase transitions between them 
\cite{chen:2013,duivenvoorden:2013,bachmann:2015a}.

Progress in the classification of ground state phases and in our understanding 
of topologically order in many-body models has mostly come from the study
of classes of models with simplifying features. E.g, Kitaev's Toric Code model
\cite{kitaev:2003}
and the Levin-Wen models \cite{levin:2005} are frustration-free and have the additional 
property that the Hamiltonian is a sum of commuting terms. Models with Matrix 
Product Ground States (MPS) in one dimension \cite{fannes:1992,verstraete:2008} or Projected
Entangled Pair States (PEPS) in two and more dimensions \cite{verstraete:2004,perez-garcia:2010,cirac:2013} 
have also been studied with considerable success.
Isolating the features
of interest in models with ground states that are as simple as possible has proved
to be a productive strategy. The present work follows the same philosophy.

The Product Vacua with Boundary States (PVBS) models, introduced
in one dimension in \cite{bachmann:2012}, were generalized to $d$
dimensions in \cite{bachmann:2015}. In the latter work the authors proved
in a particular example that a model that has a non-vanishing spectral gap
above the ground state in the thermodynamic limit and when defined on 
finite rectangular boxes, may have a spectral gap that tends to zero when 
defined on a sequence of diamond-shape finite volumes (with edges at 
45 degree angles between the edges and the coordinate axes) of increasing 
size. It was shown that the GNS Hamiltonian for the model on such an infinite
half-space is gapless due to edge excitations while it remains gapped in the bulk.

In this work we extend the results of \cite{bachmann:2015} by considering 
the spin-$1/2$ model (corresponding to a single species of particles) on 
half-spaces in $\Ir^d$ bounded by an arbitrary hyperplane. The Hamiltonian
has XXZ-type nearest neighbor interactions depending 
on parameters $\lambda_j\in (0,\infty)$, one for each coordinate direction, see
\eq{hamiltonian} for the definition. For any given values of the parameters, we prove an upper bound for the
spectral gap above the unique ground state of these models, which vanishes
for exactly one direction of the normal vector $m$. For all other choices of $m$
we derive a positive lower bound of the spectral gap except for the case $\lambda_1
=\cdots =\lambda_d=1$, which is known to have gapless excitations in the bulk.
The gapless situation, while occurring only for one specific orientation of 
bounding hyperplane, is nevertheless of particular interest. It is possible that
an entropic selection effect occurs that makes interfaces and free surfaces in 
the gapless direction more common than one might {\em a priori} expect.
Regardless, the mathematical and physical properties of the gapless 
edge spectrum deserve to be explored more fully in future work. For instance, 
PVBS models with gapless edges states in the presence of disorder would 
provide an interesting framework to study Many-Body Localization effects
at interfaces.

\section{Definition of the Model and Main Results} \label{sec:TheModel}

\subsection{The single species PVBS Model}
In this paper we consider the single species PVBS model on $\Ir^d$ as introduced
in \cite{bachmann:2015}, and adopt the notations of that paper. Let $e_1,\ldots,
e_d$ denote the canonical basis vectors of $\Ir^d$ and, by the natural
embedding, also of $\Rl^d$. For a finite lattice $\Lambda\subseteq \bZ^d$, we
associate with each site $x\in\Lambda$ the two-dimensional Hilbert space $\caH_x
= \bC^2$, with the orthonormal basis vectors $\ket{0}$ and $\ket{1}$ describing
a site that is either empty or occupied by a particle. The Hilbert space for the
system on $\Lambda$ is given by the tensor product $\cH_{\Lambda} =
\otimes_{x\in\Lambda}\cH_{x}$. The Hamiltonian $H^\Lambda$ is a sum of
projections $h_{x, x+e_j}$,with $j=1,\ldots, d$, such that $x, x+e_j\in\Lambda$.
The nearest neighbor interactions act non-trivially only on two copies of
$\Cx^2$ and depend on parameters $\lambda_j \in (0,\infty)$, $j = 1, \ldots, \,
d$. They are defined by
\be\label{interaction}
h_{x, x+e_j}= \ketbra{\phi^{(\lambda_j)}} + \ketbra{11},
\ee
where $\phi^{(\lambda)} = (\ket{01}-\lambda\ket{10})/\sqrt{1 + \lambda^2}$, for $\lambda\in (0,\infty)$. 
Here $\ket{01}$ is shorthand for $\ket{0}\otimes\ket{1}$ where the first tensor factor is associated with 
the site $x$ and the second with $x+e_j$, etc.
The Hamiltonian is defined by
\be\label{hamiltonian}
	H^\Lambda = \sum_{j=1}^d\sum_{x\in\Lambda\atop {\rm s.t.\ } x+e_j \in \Lambda }
	h_{x, x+e_j}
\ee
which is frustration-free and translation invariant.  Alternative expressions for this Hamiltonian in terms 
of spin matrices and hard-core Boson creation and annihilation operators are given in
\cite{bachmann:2015}.

As shown in
~\cite{bachmann:2015}, for a bounded and connected volume $\Lambda$, this
Hamiltonian has a two-dimensional ground state space which is spanned by the
zero particle state and a one-particle state given as follows:
\be\label{groundstates}
	\Psi_0^\Lambda = |0\rangle^\Lambda
	:=\otimes_{x\in\Lambda}|0\rangle\, , \quad\Psi_1^\Lambda 
	= \frac{1}{\sqrt{C(\Lambda)}}\sum_{x \in \Lambda}\lambda^{x} \sigma^1_x |0\rangle^\Lambda 
\ee
where $\sigma^1_x$ is the first Pauli matrix acting on the site $x$, and
$\sigma^1_x |0\rangle^\Lambda$ is the state with one particle at
site $x$. Here and in the rest of the paper, $\lambda^{x} = \prod_{j=1}^d
\lambda_j^{x_j}$. $C(\Lambda)$ is the normalization factor so that $\Vert
\Psi_1^\Lambda\Vert=1$.
It is given by
\be\label{CLambda}
C(\Lambda) = \sum_{x\in\Lambda} \lambda^{2x}.
\ee

We are interested in the excitation spectrum of this model defined on infinite
half spaces bounded by a hyperplane containing the origin, that is subsets
$D\subset \Ir^d$ determined by a unit vector $m\in\Rl^d$ (the inward normal) as
follows:
\begin{align}\label{HyperplaneVolume}
	D:=\{x \in \mathbb{Z}^d: m \cdot x \geq 0\}.
\end{align}
More precisely, we are interested in the spectrum of the Hamiltonian $H^D$ in
the GNS representation of the zero particle ground state on $D$. Since this
ground state is given by a tensor product vector, the GNS representation can be
given explicitly. The GNS Hilbert space is generated by all states with only
a finite number of occupied sites. The dense subspace spanned by all such
vectors is a core for the self-adjoint operator $H^D$. As detailed in
\cite{bachmann:2015}, $H^D$ is non-negative, and has a one-dimensional kernel
spanned by the ground state $\Psi_0^D$.

In this work, the main question we address is the existence or nonexistence of a
non-vanishing spectral gap above the ground state. For any Hamiltonian $H$,
such that $H\geq 0$ and $0\in\spec (H)$, the spectral gap $\gamma(H)$ is
defined as follows:
\be\label{gap|}
	\gamma(H) := \sup\{\delta>0: \spec(H)\cap(0,\delta) = \emptyset  \}
\ee
with the convention that $\gamma(H) =0$ if the set on the RHS is empty.
In the latter case we call the model gapless. We will often denote $\gamma(H^\Lambda)$ 
by $\gamma(\Lambda)$ and $\gamma(H^D)$ by $\gamma(D)$ to simplify the notation.

\subsection{Summary of Results}

We recall that the domain $D$ depends on a unit vector $m\in\Rl^d$, and that the
Hamiltonian $H^D$ additionally depends on a vector of parameters $\vec\lambda =
(\lambda_1, \ldots, \lambda_d)\in (0,\infty)^d$. We will often use the notation
$\log\vec\lambda$ for the vector $(\log\lambda_1, \ldots, \log \lambda_d)$. The
goal of this work is to determine for each combination of unit inward pointing
normal $m\in\Rl^d$ and parameters $\vec{\lambda}\in(0,\infty)^d$ whether the gap
$\gamma(H^D)$ vanishes or not and to find explicit bounds for the gap whenever
possible. The case $d=1$ was treated in \cite{bachmann:2012} where it was shown
that model is gapless if $\lambda =1$ and gapped otherwise. Special for the case
$d=1$ is that the ground state space for the half-infinite chain
$[1,\infty)\subset \Ir$ is two-dimensional if $m\log \lambda<0$ and
one-dimensional if $m\log \lambda\geq 0$. From now on we will only discuss
dimensions $d\geq 2$, in which case the ground state space for any half-space is
one-dimensional.

For $d\geq 2$, the positivity of $\gamma(D)$ is determined by the angle,
$\theta$, between the vectors $m$ and $-\log \vec\lambda$, which is well defined 
except in the case $\vec\lambda=(1,\ldots, 1)$. It was already proved in 
\cite{bachmann:2015} that the model is gapless for $\vec\lambda=(1,\ldots, 1)$
and is not considered here.

Our first result is an upper bound. For its statement, we define
\be\label{cv}
c(v) := \min\{|v_j|:v_j \neq 0\},  v\in \Rl^d.
\ee

\begin{theorem}[Upper Bound]\label{thm:upperbound}
For all $d\geq 2$, $\lambda_1,\ldots,\lambda_d\in (0,\infty)$, and unit vectors $m\in\Rl^d$ such that $m\cdot\log\vec\lambda <0$,
one has the following upper bound:
\be\label{upperbound}
	\gamma(H^D) \leq \frac{2(d-1)}{c(m) c(\vec\lambda)^2}\|\log\vec{\lambda}\||\sin(\theta)|,
\ee
where $\theta$ is the angle between the vectors $-m$ and $\log \vec\lambda$. In particular, the gap vanishes if  $\theta=0$.
\end{theorem}
The vanishing of the gap at $\theta=0$ is due to the appearance of extended edge states. The probability distribution for the
position of the particle in the one-particle ground state is proportional to $\lambda^{2x}$. It is easy to see that,
if $\log \vec\lambda$ is an outward normal to the boundary, the one-particle ground state
assigns approximately equal probability for the position of the particle everywhere along the boundary,
and the probability decays exponentially in the distance from the boundary.

The following theorem shows that whenever the upper bound given in \eq{upperbound} does not
vanish, the model does in fact have a non-vanishing spectral gap.

\begin{theorem}[Existence of a Spectral Gap]\label{thm:Gapped}
If $\log\vec\lambda \neq -\Vert\log\vec\lambda\Vert m$, then $\gamma(H^D) >0 $.
\end{theorem}

This theorem states that if the angle between $\log\vec{\lambda}$ and $-m$
is nonzero, then $H^D$ is gapped. 

\subsection{About the Proofs}
The results in this work are proved by deriving upper and lower bounds on the
spectral gap. Theorem \ref{thm:upperbound} is proved by a variational
calculation in Section \ref{sec:upperbound}.
For the proof of Theorem \ref{thm:Gapped}, we use a well-known relationship
between the infinite volume spectral gaps and finite volume spectral gaps as
well as the martingale method, all of which we described in detail in Section
\ref{sec:MartingaleMethod}.

Our emphasis in proving upper and lower bounds is to establish in all cases whether there is a positive
spectral gap in the thermodynamic limit or not. We have not attempted to obtain best possible bounds.
For instance, in the gapless cases we have not established the rate with which the finite volume 
lowest energy excitations vanish in the thermodynamic limit. The gap vanishes at least as fast
as $O(L^{-1})$, where $L$ is the diameter of the support of the excitation. In some 
cases we can show an upper bound of the form $O(L^{-2})$. We have not shown that
the latter behavior holds in general.

Since the martingale method proves lower bounds for spectral gaps for finite volume Hamiltonians
and the PVBS models are translation invariant, the following corollary
immediately follows from our proof of Theorem \ref{thm:Gapped}.

\begin{corollary}\label{cor:GappedZd}
The PVBS model with one species of particle on $\bZ^d$ and model parameters
$\lambda_1, \ldots, \, \lambda_d \in (0, \, \infty)$ is gapped if there exists
at least one $j$ such that $\lambda_j \neq 1$.
\end{corollary}

It was shown in \cite{bachmann:2015} that the PVBS model on $\bZ^d$ is
gapless if all $\lambda_j = 1$. Therefore, Corollary \ref{cor:GappedZd}
completes the gap classification for the one species PVBS models on $\bZ^d$.

The proof of Theorem \ref{thm:Gapped} for a general dimension $d$ is more easily
understood by first considering the case of $d=2$. For this reason, we prove in
detail Theorem \ref{thm:Gapped} for $d=2$ in Section \ref{sec:TwoDimCase}, and
refer to it as necessary when proving the general result in Section
\ref{sec:dDimCase}. Furthermore, in addition to the statement of Theorem
\ref{thm:Gapped}, we also prove explicit lower bounds for the well-chosen finite
volumes and, as a consequence, we obtain lower bounds for the gap in the
thermodynamic limit. The dependence of these bounds on the parameters $m$ and
$\vec \lambda$ is somewhat involved, however, which is why we did not include
the bounds in the statement of Theorem \ref{thm:Gapped}. They can be found in
Sections \ref{sec:TwoDimCase} and \ref{sec:dDimCase}.

To simplify the proofs, we will permute and reflect coordinates in $\Ir^d$ such that 
the components of the inward unit normal of the half-space under consideration 
are non-negative and the first component is non-zero. That this can be done
without loss of generality is easy to see based on the following observations.

First, replacing $m_j$ by $-m_j$, corresponds to a reflection of $\Lambda$
through the hyperplane normal to the $j$th basis vector $e_j$. The only terms in
the Hamiltonian affected by this reflection are the interactions of nearest
neighbor pairs of the form $(x,x+e_j)$. Let $R$ denote the unitary interchanging
the tensor factors in $\cH_x\otimes \cH_{x+e_j}$. Then
$R\phi^{(\lambda)}=\phi^{(\lambda^{-1})}$ and it follows that $R h^{(\lambda)}
R^*=h^{(\lambda^{-1})}$. So, we can make all components of $m$ non-negative by
replacing some of the parameters $\lambda_j$ with $\lambda_j^{-1}$.

Second, since we can assume that all components of $m$ are non-negative and
since $m\neq 0$, at least one of the components is strictly positive. We can
therefore relabel the coordinates such that $m_1>0$. Such a relabeling
corresponds to a permutation of the parameters $\lambda_j$.

For future reference we summarize these observations in the following remark.

\begin{remark}\label{rem:coordinates}
The model \eq{hamiltonian} with parameters $\lambda_1,\ldots,\lambda_d\in (0,\infty)$, 
defined on finite subsets $\Lambda$ of the half-space $D\subset \Ir^d$,
bounded by the hyperplane containing the origin with inward normal 
$m=(m_1,\ldots,m_d)\in \Rl^d$, is unitarily equivalent to the model on a finite
volume $\tilde\Lambda$ with parameters 
$\tilde\lambda_1,\ldots,\tilde\lambda_d\in (0,\infty)$ and normal vector 
$\tilde m$ such that $\tilde m_1>0$, and $\tilde m_2\geq 0,
\dots, \tilde m_d\geq 0$.
\end{remark}


\section{Iterative Method for Proving Lower Bounds for Spectral Gaps}

\subsection{The Martingale Method}\label{sec:MartingaleMethod}
To prove that there is a non-vanishing spectral gap above the ground state in the
thermodynamic limit we rely on the following well-known theorem to reduce the 
problem to finding lower bounds for the gaps for a suitable family of finite
volumes.

\begin{theorem}\label{thm:GNSSpecGap}
Let $H^D$ be the GNS Hamiltonian associated with the connected infinite volume
$D$ with spectral gap $\gamma(D)$. Then for any sequence of increasing and
absorbing volumes $\Lambda_L \nearrow D$,
\begin{align*}
\gamma(D) \geq \limsup \gamma(\Lambda_L),
\end{align*}
where $\gamma(\Lambda_L)$ is the spectral gap of the frustration-free
Hamiltonian $H^{\Lambda_L}$.
\end{theorem}

To estimate finite volume gaps, we use an approach called the martingale
method which is given in the following theorem. It provides
conditions under which the spectral gap for a frustration-free model on a finite
volume $\Lambda$, such as the PVBS model considered here, can be bounded by a
fraction of the gap for the model on small subvolumes.

\begin{theorem}[Martingale Method, \cite{nachtergaele1996spectral}]\label{thm:MartingaleMethod}
For a finite volume $\Lambda$ and frustration-free Hamiltonian $H^\Lambda$ let
$\Lambda_n$ be a finite sequence of volumes with $\Lambda_0 = \emptyset$ and
$\Lambda_n \nearrow\Lambda_L=\Lambda$ such that the following three conditions
hold for the local Hamiltonians for the same $\ell\geq 2$:\\
(i) For some positive constant $d_\ell$,
\begin{equation*}
\sum_{n=\ell}^L H^{\Lambda_n\backslash\Lambda_{n-\ell}} \leq d_\ell H^{\Lambda_L}
\end{equation*}
(ii) For some positive constant $\gamma_\ell$ and $n_\ell$, if $n\geq n_\ell$,
\begin{equation*}
H^{\Lambda_n\backslash\Lambda_{n-\ell}} \geq \gamma_\ell(\mathbb{I} -
G^{\Lambda_n\backslash\Lambda_{n-\ell}})
\end{equation*}
where $G^{\Lambda}$ is the orthogonal projection onto $\caG_\Lambda =
\ker(H^\Lambda)$. 
\vskip 12pt
\noindent
(iii) There exists a constant $\epsilon_\ell <
\frac{1}{\sqrt{\ell}}$ and $n_\ell$ such that $n_\ell\leq n\leq L-1$
\begin{equation*}
\|G^{\Lambda_{n+1}\backslash\Lambda_{n+1-\ell}}E_n\| \leq \epsilon_\ell
\end{equation*}
where $E_n := G^{\Lambda_n} - G^{\Lambda_{n+1}}$ is the projection onto
$\caG_{\Lambda_n}\cap \caG^\perp_{\Lambda_{n+1}}$.

Then the spectral gap of $H^{\Lambda_L}$ satisfies
\begin{equation*}
\gamma(\Lambda_L) \geq \frac{\gamma_\ell}{d_\ell}(1-\epsilon_\ell
\sqrt{\ell})^2.
\end{equation*}
\end{theorem}

In a typical application to quantum spin chains, the sequence $(\Lambda_n)$ is
simply a sequence of increasing intervals, say $\Lambda_n=[1,n]$, and the
conditions are satisfied with $d_\ell=\ell$, and $\ell=2$. Spin ladder systems
can be treated in the same way with $\Lambda_n=[1,n]\times [1,2]\subset \Ir^2$. In \cite{bachmann:2015} we applied
the method to PVBS models defined on $d$-dimensional boxes of the form
$[1,L_1]\times\cdots \times [1,L_d]$, by using induction on the dimension. For
example, in the case that $d=2$, we apply Theorem \ref{thm:MartingaleMethod} to
the sequence $\Lambda_n= [1,L_1]\times [1,n]$, $n=2,\ldots L_2$. Condition (ii),
with $\ell=2$, is shown to be satisfied by applying the theorem to the spin
ladder $\Lambda_{n}\setminus \Lambda_{n-2}=[1,L_1]\times [n-1,n]$. Due to the
translation invariance of the model this yields a lower bound for the gap for
the spin ladders independent of $n$. Applying the martingale method to the
spin ladder provides a lower bound of the spectral gap independent of $L_1$ and
$L_2$. This reasoning can be repeated to estimate the gap for $d$-dimensional
boxes.
The result is a lower bound of the general form
\be\label{boundforboxes}
\gamma(\Ir^d)\geq 2^{-d}\gamma(B_d)(1-\sqrt{2}\epsilon_1)^2 \cdots
(1-\sqrt{2}\epsilon_d)^2,
\ee
where $B_d$ is the $d$-dimensional unit hypercube.

In the situation at hand, we will prove a lower bound on the gap for a sequence
of finite volumes $\Lambda_L$ that increases to the half-space $D$ as
$L\to\infty$. Unless $m=e_j$ for some $j=1,\ldots,d$, $\Lambda_L$ cannot be
rectangular boxes, i.e., a cartesian product of intervals. We will choose
volumes $\Lambda_L$ that are bounded by $2d$ hyperplanes, one of which aligns
with the bounding hyperplane of $D$. In order to avoid a vanishing gap along the
sequence $\Lambda_L$, the outward normals of all bounding hyperplanes must have
a non-vanishing angle with $\log \vec\lambda$. This requirement follows from
Theorem \ref{thm:upperbound}. The size of $\Lambda_L$ will scale linearly with
$L$ in each direction.

For each $\Lambda_L$, we will construct a suitable sequence $\Lambda_n$
for which the martingale method applies. In analogy to the treatment of
rectangular boxes referred to above, we will apply the martingale method inductively as the volumes grows in one coordinate 
direction at a time. Care must be taken so that
the Conditions (i)--(iii) can be verified for each step. We discuss what this
involves for each condition.

Condition (i) is given by an upper bound on the number of subvolumes
$\Lambda_n\backslash\Lambda_{n-\ell}$ that contain the support of any given
interaction term. In this paper, it will always be clear that we can take $d_\ell =\ell$. 
For this reason, we will not refer to $d_\ell$ in our estimates.

Condition (ii) is slightly complicated by the fact that, depending on the normal
$m$, $D$ may not be translation invariant in sufficiently many directions.
It is therefore not possible, in general, to choose sequences $\Lambda_n$ such
that $\Lambda_n \setminus \Lambda_{n-\ell}$ are isomorphic for all values of
$n$, as was the case for rectangular boxes. As a consequence, the quantity
$\gamma_\ell$ for Condition (ii) will be the minimum over a finite number of
finite-volume gaps. 

In order to satisfy Condition (iii), the volumes $\Lambda_n
\setminus \Lambda_{n-\ell}$ should be connected (in the nearest neighbor sense).
As demonstrated in Figure \ref{fig:Connectedness}, this requires that we
consider larger values of $\ell$ due to the slanted boundaries, and this value
may be different for each application of the martingale method. It is important,
however, that we will need only a finite number of different values of $\ell$,
i.e., the $d$ application of Theorem \ref{thm:MartingaleMethod} can be achieved
with bounded $\ell_1,\ldots, \ell_d$, uniformly in $L$. This leads to lower
bounds of the form
\be\label{generallowerbound}
\gamma(\Lambda_L) \geq  \min_{\Lambda\in
				  \mathcal{V}(\vec{\ell})}\gamma(\Lambda)\prod_{j=1}^d
				  \frac{(1-\epsilon_{\ell_j}\sqrt{\ell_j})^2}{\ell_j} 
\ee
where $\mathcal{V}(\vec{\ell})$ is a finite collection of volumes
for which the length in the $j$-th coordinate direction of any $\Lambda\in
\mathcal{V}(\vec{\ell})$ is proportional to $\ell_j$. Furthermore, the set
$\mathcal{V}(\vec{\ell})$ is the same set for the lower bound estimate for every
volume $\Lambda_L$.

In general, the most difficult step in applying the martingale method is finding
a sequence of volumes which satisfies Condition (iii). However, when applying
the martingale method for the PVBS models on a sequence of connected lattices
such that each subvolume $\Lambda_n\backslash\Lambda_{n-\ell}$ is also
connected, the operator norm in Condition (iii) can be exactly calculated and is
written in terms of the normalization coefficients for the one particle ground
states on several volumes. This result is given in the lemma below, and the
proof is given in the Appendix.

\begin{lemma} \label{lem:EpsilonCalculation}
For each $\ell\geq 2$, and a sequence of increasing finite volumes
$\Lambda_n\nearrow \Lambda$, $n\geq 1$, such that $\Lambda_n$ and
$\Lambda_{n}\backslash\Lambda_{n-\ell}$ are connected for all $n$, the operator
norm in Condition (iii) of Theorem \ref{thm:MartingaleMethod} applied to the
PVBS model \eq{hamiltonian} is given by
\begin{equation}\label{epsilon}
\|G^{\Lambda_{n+1}\backslash\Lambda_{n+1-\ell}}E_n\|^2 =
\frac{C(\Lambda_{n+1-\ell})C(\Lambda_{n+1}\backslash\Lambda_n)}{C(\Lambda_n)
C(\Lambda_{n+1}\backslash\Lambda_{n+1-\ell})},
\end{equation}
where $C(\Lambda)$ denotes the normalization coefficient defined in \eq{CLambda}.
\end{lemma}
It is remarkable that the operator norm can be exactly calculated. This lemma will be applied
to a variety of situations. In each case we will find a simple upper bound for the RHS of
\eq{epsilon}, depending on the geometry of the volumes (i.e., $\Lambda_n$), and
$\ell$.

\subsection{Choosing Volumes} \label{sec:VolumeChoices}
From the above discussion it is clear that we need to make judicious choices for
the sequences of finite volumes appearing the proofs. As we have already
discussed, the following three properties are necessary for the finite
volumes $\Lambda_n$ and $\Lambda_n\setminus\Lambda_{n-\ell}$. First, the
boundary of $\Lambda_n$ has to include an increasingly large subset of the
boundary of $D$ (i.e., the hyperplane with inward normal $m$). Second, the
orientation of large boundary surfaces should be such that the angle between
their outward normal and the vector $\log\vec\lambda$ remains bounded away from
zero. Third, they should be connected. However, the satisfaction Condition (iii)
is also dependent on choosing volumes that fulfill the following heuristic.

The general heuristic for selecting a sequence of volumes such that Condition (iii) holds is to choose the
volumes where the magnitude of the single particle ground state at site 
$x$, $\lambda^{2x} = \exp(2x\cdot\log\vec{\lambda})$,
is maximized at a unique point in each volume when considered as a subset of $\mathbb{R}^d$.
The maxima occur on the boundary of the convex volumes at the points furthest in the
$\log\vec{\lambda}$ direction.
The uniqueness of the maximum implies that the maxima over $\Lambda_{n+1}\setminus\Lambda_n$ 
and $\Lambda_n\setminus\Lambda_{n-1}$ are not equal
because there is a unique maximum in $\Lambda_{n+1}\setminus\Lambda_{n-1}$.
Due to the exponential form of the 1-particle ground state wave function, the maxima differ approximately 
by an factor not equal to one and consequently the normalization constants for $\Lambda_{n+1}\setminus\Lambda_n$ and $\Lambda_{n}\setminus\Lambda_{n-1}$ differ by this factor.
It follows that the normalization constants in the RHS of \eq{epsilon} are approximately equal to a ratio of 
geometric sums that decays, which is sufficient to satisfy Condition (iii).

If there is not a unique maximum, Condition (iii) fails.  In this case,  maxima differ by a factor approximately equal to one and the RHS of \eq{epsilon} are not geometric sums but sums of ones.  It follows that the operator norm of
Condition (iii) will converge $1/\sqrt{\ell}$ as $n\to\infty$ and the martingale
method will not give a strictly positive lower bound.
In this case, the set of maximizing points form a line-segment or bounded
hyperplane spanned by vectors perpendicular to both $\log\vec{\lambda}$ and the
normal vectors to the boundary hyperplanes which contain a maximum.
As the volumes grow as $L\to\infty$, these sets of maxima can support extended
states which in the finite volumes have energy converging to zero.
These low energy extended states are an artifact of the choice of finite volumes
in the sense that they do not correspond to low-energy states in the infinite
system we are considering. By choosing volumes which have a unique maximum, we
avoid this issue.


\section{Existence of a Spectral Gap in Two Dimensions}\label{sec:TwoDimCase}

\begin{theorem}[Existence of a Spectral Gap for $d=2$]\label{thm:TwoDimGapped}
Let $D :=\{x\in\mathbb{Z}^2: m\cdot x \geq 0 \}$.  If
$$
(\log\lambda_1,\log\lambda_2) \neq -\|\log\vec{\lambda}\|(m_1,m_2),
$$
then $H^D$ is gapped.
\end{theorem}

We give a detailed proof of applying the martingale method to $H^D$ where
$D\subseteq\mathbb{Z}^2$ to motivate the proofs in higher dimensions. Thus, we
choose a sequence of finite volumes $\Lambda_L$ that increase to $D$ such that
the martingale method proves a nonzero lower bound for the spectral gap of each
Hamiltonian $H^{\Lambda_L}$ which is uniform in $L$, and appeal to Theorem
\ref{thm:GNSSpecGap} to show the spectral gap $\gamma(D)$ also satisfies the
same lower bound.

As discussed in Section \ref{sec:VolumeChoices}, when applying the martingale method
to PVBS models with slant boundaries, it is essential to make good choices for the geometry 
of the finite volumes. Let $\theta$ be the angle between $-m$ and
$\log\vec{\lambda}$. For $d=2$ and $\theta\neq 0$, the general heuristic from
Section \ref{sec:VolumeChoices} is satisfied for any sequence of volumes for
which $\log\vec{\lambda}$ is not an outward pointing normal to any boundary. 

Using the symmetries of the PVBS model outlined in Remark \ref{rem:coordinates},
we prove the gapped statement for $d=2$ by considering two types of finite
volume sequences. In the case that $\theta \neq \pi$, we use parallelograms with
a pair of boundaries parallel to the boundary of $D$. We further divide this
into two subcases based on the inward normal $m$. Case 1a is the case where
there is a $j$ such that $\lambda_j \neq 1$ and $m_j \neq 0$.  Case 1b is the
case where such a $j$ does not exist. The choice of paralellograms fails if
$\theta = \pi$ as then $\log\vec{\lambda}$ is the outward pointing normal to the
side of the parallelogram opposite to the boundary of $D$. In this case, we
modify the parallelogram to a trapezoid. We once again break this into two
subcases. Case 2a is the case that both $m_1$ and $m_2$ are not equal to zero.
Case 2b is the case where either $m_1$ or $m_2$ is zero, but not both as $m$ is
a unit vector.

The process of applying the martingale method is the same strategy for all four
cases. We first define the sequence of finite volumes $\Lambda\nearrow D$. 
We then bound the normalization coefficient $C(\Lambda)$ for a general finite
volume of the geometry we have chosen, which we use when applying Lemma
\ref{lem:EpsilonCalculation}. Lemma \ref{lem:EpsilonCalculation} only applies
the finite volumes are connected as subgraphs of $\bZ^2$. This condition
holds for all volumes we choose except for in Case 1a. There, we will determine
a sufficient condition for connectedness. In all other cases, we will forgo any
additional comments on connectedness.

For each case, we apply the martingale method twice. After the first application
(in the direction of $x_2$), we will need to estimate the gap for a sequence of
quasi-one dimensional systems obtained from the original sequence of finite
volumes. This is done by a second application (in the direction of $x_1$)
and yields a uniform lower bound. The following function $f$ will often appear
in the calculation of $\epsilon_\ell$ for Condition (iii):
\begin{equation*}
f(n,\ell) =
\frac{\lambda^{2(\ell-1)}(1-\lambda^{2(n+1-\ell)})(1-\lambda^2)}{(1-\lambda^{2\ell})(1-\lambda^{2n})}
\end{equation*}
Specifically, there will be some positive constant $C$ for which we find
estimates of the form
\begin{equation}\label{CommonBound}
\|G^{\Lambda_{n+1}\backslash\Lambda_{n+1-\ell}}E_n\|^2\leq C\cdot f(n,\ell).
\end{equation}
The function $f(n,\ell)$ is increasing in $n$.
By treating $\lambda<1$ and $\lambda>1$ separately and taking 
the limit $n\to\infty$ it follows that $f(n,\ell) \leq \min(1,
\lambda^{2(\ell-1)})\cdot \frac{1-\lambda^2}{1-\lambda^{2\ell}},$ which decays
exponentially in $\ell.$ Hence, there exists a minimal $\ell$ that depends on
$C$ such that
\begin{equation*}
C\cdot \frac{(1-\lambda^2)\min(1, \lambda^{2(\ell-1)})}{1-\lambda^{2\ell}}
	< \frac{1}{\ell}.
\end{equation*}
In these situation, Condition (iii) of the martingale method is satisfied for
\begin{equation}\label{CommonEpsilon}
\epsilon_\ell = \sqrt{C}\min(1,\,
\lambda^{\ell-1})\sqrt{\frac{1-\lambda^2}{1-\lambda^{2\ell}}}
\end{equation}
for smallest value $\ell$. In any situation where we find an upper bound of
the form \eqref{CommonBound}, we will immediately use this definition for
$\epsilon_\ell$.

\begin{figure}[t]
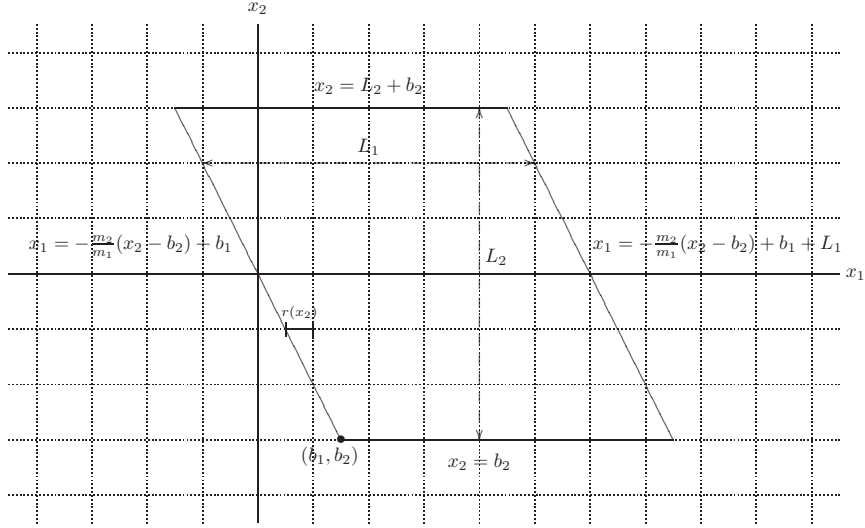

\begin{adjustbox}{max totalsize={.7\textwidth}{.7\textheight}}
{
\xy
(0,-25); (0,65) ** \dir{-}; 
(0,68)*+{x_2};
(-45,20); (105,20) ** \dir{-}; 
(108,20)*+{x_1};
(-40,-25);(-40,65) **\dir{.}; 
(-30,-25);(-30,65) **\dir{.};
(-20,-25);(-20,65) **\dir{.};
(-10,-25);(-10,65) **\dir{.};
(10,-25);(10,65) **\dir{.};
(20,-25);(20,65) **\dir{.};
(30,-25);(30,65) **\dir{.};
(40,-25);(40,65) **\dir{.};
(50,-25);(50,65) **\dir{.};
(60,-25);(60,65) **\dir{.};
(70,-25);(70,65) **\dir{.};
(80,-25);(80,65) **\dir{.};
(90,-25);(90,65) **\dir{.};
(100,-25);(100,65) **\dir{.};
(-45,-20); (105,-20) ** \dir{.}; 
(-45,-10); (105,-10) ** \dir{.}; 
(-45,10); (105,10) ** \dir{.};
(-45,0); (105,0) ** \dir{.};
(-45,30); (105,30) ** \dir{.};
(-45,40); (105,40) ** \dir{.};
(-45,50); (105,50) ** \dir{.};
(-45,60); (105,60) ** \dir{.};
(-15,50); (15,-10)**\dir{-}; 
(45,50); (75,-10)**\dir{-};
(-15,50); (45,50)**\dir{-};
(15,-10); (75,-10)**\dir{-};
(83,25)*+{x_1=-\frac{m_2}{m_1}(x_2-b_2)+b_1+L_1};
(-23,25)*+{x_1=-\frac{m_2}{m_1}(x_2-b_2)+b_1};
(20,54)*+{x_2 = L_2+b_2};
(40,-14)*+{x_2 = b_2};
(15,-10)*+{\bullet};
(13,-13)*+{(b_1, b_2)};
(5,10); (10,10)**\dir{-};
(5,8.75); (5,11.25)**\dir{-};
(10,8.75); (10,11.25)**\dir{-};
(7.5,13)*+{\scriptstyle r(x_2)};
(40,-10); (40,50)**\dir{--}; 
{\ar@{->} (40,49.9)*{}; (40,50)*{}};
{\ar@{->} (40,-9.9)*{}; (40,-10)*{}};
(43,23)*+{L_2};
(-10,40); (50,40)**\dir{--}; 
{\ar@{->} (49.9,40)*{}; (50,40)*{}};
{\ar@{->} (-9.9,40)*{}; (-10,40)*{}};
(20,43)*+{L_1};
\endxy 
}
\end{adjustbox}
   \caption{A general parallelogram $P(b_1, b_2, L_1, L_2)$.}
   \label{fig:parallelogram}
\end{figure}

\vskip12pt
\noindent
\textbf{Case 1a}:  Suppose that $(\log \lambda_1, \log \lambda_2) \neq
\|\log\vec{\lambda}\|(m_1,m_2)$ and there exists a $j$ such that $\lambda_j \neq
1$ and $m_j \neq 0$. Without loss of generality, permute the indices so
$\lambda_1 \neq 1$ and $m_1\neq 0$.
\vskip12pt
\textbf{Volumes:} 
The sequence of volumes $\Lambda_L\nearrow D$ we choose are parallelograms that
align with the boundary of $D$. Other parallelograms need to be considered 
to verify Condition (ii) of the martingale method so we define a general
parallelogram $P(b_1,b_2,L_1,L_2)$ using a base point, $b = (b_1, \, b_2)$, and
length parameters $L_1,$ $L_2$. Specifically,
\begin{align*}
P(b_1,b_2,L_1,L_2) =\{x\in\mathbb{Z}^2: 0\leq \frac{m}{m_1} \cdot
(x-b) < L_1,\ 0 \leq x_2 -b_2 < L_2 \}.
\end{align*}
See Figure \ref{fig:parallelogram}.
The sequence of volumes $\Lambda_L\nearrow D$ is given by
$\Lambda_L = P(\frac{m_2}{m_1}L, \, -L, \, 2L, \, 2L)$.

\begin{figure}[h]
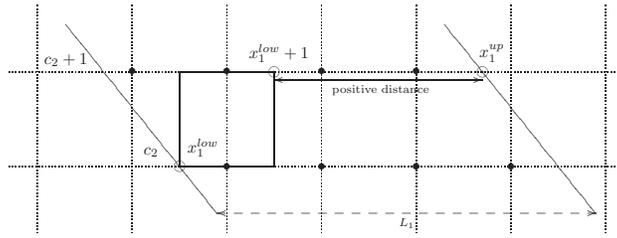

\begin{adjustbox}{max totalsize={.5\textwidth}{.7\textheight}}
{
\xy
(0,-10);  (-32,30) **\dir{-};
{\ar@{<-->}_{L_1} (0,-10)*{}; (80,-10)*{}};
(80,-10);(48,30)**\dir{-};
(-44,0);(86,0)**\dir{.};
(-14,3)*+{c_2};
(-44,20);(86,20)**\dir{.};
(-32,22.5)*+{c_2+1};
(-38,-14);(-38,34)**\dir{.};
(-18,-14);(-18,34)**\dir{.};
(2,-14);(2,34)**\dir{.};
(22,-14);(22,34)**\dir{.};
(42,-14);(42,34)**\dir{.};
(62,-14);(62,34)**\dir{.};
(82,-14);(82,34)**\dir{.};
(2,0)*+{\bullet};
(22,0)*+{\bullet};
(42,0)*+{\bullet};
(62,0)*+{\bullet};
(-18,20)*+{\bullet};
(2,20)*+{\bullet};
(22,20)*+{\bullet};
(42,20)*+{\bullet};
{\ar@{<->}_{\text{ positive distance}} (12,18.25)*{}; (56,18.25)*{}};
(12,20)*+{\odot};
(13,24)*+{x_1^{low}+1};
(56,20)*+{\odot};
(58,24)*+{x_1^{up}};
(-8,0)*+{\odot};
(-3,4)*+{x_1^{low}};
(-8,0);(-8,20)**\dir{-};
(-8,20);(12,20)**\dir{-};
(-8,0);(12,0)**\dir{-};
(12,0);(12,20)**\dir{-};
\endxy 
}
\end{adjustbox}
   \caption{Diagram for connectedness.} 
  \label{fig:Connectedness}
\end{figure}

To apply Lemma \ref{lem:EpsilonCalculation}, we need to ensure our
volumes are connected. We prove that the condition $L_1 \geq\frac{m_2}{m_1}+1$
is sufficient to guarantee the connectedness of $P(b_1, \, b_2, \, L_1, \, L_2)$.
Let $x_2 = c_2$ and $x_2=c_2+1$ be two subsequent values between
$b_2$ and $L_2+b_2$. The parallelogram $P(b_1, b_2, L_1, L_2)$ is connected if and only if
there is a $x_1=c_1$ such that $(c_1,c_2), \, (c_1, c_2+1)\in
P(b_1,b_2,L_1,L_2)$. To guarantee the existence of such a $c_1$, it is
sufficient to show
\begin{equation}\label{Connectedness}
x_1^{low}(c_2)+1\leq x_1^{up}(c_2+1)
\end{equation}
where $x_1^{low}(c_2) = -\tfrac{m_2}{m_1}(x_2-b_2)+b_1$ is the lower bound for
all $x_1$ associated with $x_2=c_2$, and  $x_1^{up}(c_2+1) = 
-\tfrac{m_2}{m_1}(x_2+1-b_2)+b_1+L_1$ is the upper bound for all $x_1$
associated with $x_2=c_2+1$. 
This is summarized in Figure \ref{fig:Connectedness}. Simplifying the inequality
in \eqref{Connectedness} shows this is equivalent to $L_1\geq \frac{m_2}{m_1}+1$.
In particular, $\Lambda_L$ is connected if $2L \geq \frac{m_2}{m_1}+1$,
and we start the sequence with a sufficiently large $L$. 

\textbf{Normalization Coefficient:}
For the parallelogram $P(b_1, \, b_2, \, L_1, \, L_2)$, we always choose $b_2, \,
L_1,$ and $L_2$ to be integer valued, while $b_1$ need only be real. The
condition $0\leq \frac{m}{m_1} \cdot (x-b) < L_1$ is treated
as a bound on $x_1$, where the smallest integer value of $x_1$ in the
paralellogram for fixed $x_2$ is 
$$x_1^{min} = \frac{-m_2}{m_1}(x_2-b_2)+b_1 +
r(x_2),$$ 
for which $r(x_2) \in [0,1)$ is a remainder term that only
depends on $x_2$, see Figure \ref{fig:parallelogram}. For a connected parallelogram $P(b_1, b_2, L_1,
L_2)$, the normalization constant is given by
 \begin{align}
C(P(b_1,b_2,L_1,L_2)) &= \sum_{x\in P(b_1,b_2,L_1,L_2)}
\lambda_1^{2x_1} \lambda_2^{2x_2} \label{eqn:ParCoeff} \\
& = \sum_{x_2=b_2}^{b_2 + L_2 -1}\lambda_2^{2x_2}\sum_{x_1=
b_1 - \frac{m_2}{m_1}(x_2 - b_2) + r(x_2)}^{b_1+L_1 -1 - \frac{m_2}{m_1}(x_2 -
b_2) + r(x_2)} \lambda_1^{2x_1} \nonumber \\
					  &= \lambda^{2b} \sum_{x_2=0}^{L_2
-1}(\lambda_1^{\frac{-m_2}{m_1}}\lambda_2)^{2x_2}
\lambda_1^{2r(x_2)}\sum_{x_1=0}^{L_1 -1} \lambda_1^{2x_1}. \nonumber
 \end{align}
\vskip12pt 
\textbf{Martingale Method in $x_2$ Direction:} Recall that the sequence
$\Lambda_L\nearrow D$ is defined by $\Lambda_L = P(\frac{m_2}{m_1}L, -L, 2L,
2L)$. Let $N = 2L$ and define $\Lambda_n^{(2)} = P(\frac{m_2}{m_1}L, -L, 2L,
n).$ Notice that $\Lambda_n^{(2)}$ only increases in the $x_2$-coordinate and
that $\Lambda_N^{(2)} = \Lambda_L$. 

We use the martingale method to find a lower bound of the spectral gap of
$H^{\Lambda_N^{(2)}}$. The martingale method allows freedom in choosing any
$\ell_2$ that satisfies the necessary conditions for Theorem
\ref{thm:MartingaleMethod}. In general, there will be an infinite number of such
$\ell_2$ that satisfy the conditions of the martingale method. However, we
always choose the lowest possible value of $\ell_2$.

As mentioned in \ref{sec:VolumeChoices}, Condition (i) is satisfied with
$d_{\ell_2} = \ell_2$. For Condition (ii), it suffices to define
$\gamma_{\ell_2}$ as
\begin{align*}
	\gamma_{\ell_2} &:= \min_{\ell_2\leq n\leq N}
	\gamma(\Lambda_n^{(2)}\backslash\Lambda_{n-\ell_2}^{(2)})\\
	& =  \min_{0\leq b_2 \leq 2L-\ell_2}
	\gamma\left(P(-\frac{m_2}{m_1}b_2,b_2,2L,\ell_2)  \right),
\end{align*}
which is greater than zero since it is a minimum of a finite collection of
spectral gaps for finite dimensional Hamiltonians. Since this gap is dependent
on $L$, it could happen that $\gamma_{\ell_2}\to 0$ as $L\to \infty$. However,
in the second application of the martingale method, we show that for all $L$,
the set of spectral gaps we minimize over to define $\gamma_{\ell_2}$ share a
common nonzero lower bound. We will apply the same reasoning and strategy for
the remaining cases.

We apply Lemma \ref{lem:EpsilonCalculation} as $2L\geq m_2/m_1+1$ guarantees
connectedness of all necessary volumes. Define $\tilde{\lambda}_2 =
\lambda_2\lambda_1^{-m_2/m_1}$. Since $\log\vec{\lambda}\neq
\pm\|\log\vec{\lambda}\|m$, we find that $\tilde{\lambda}_2 =
\lambda_2\lambda_1^{-m_2/m_1} \neq 1$. It is easily seen that for $\Lambda_k
\subseteq \Lambda_n$, 
$$C(\Lambda_n \backslash \Lambda_k) = C(\Lambda_n) -
C(\Lambda_k).$$ 
Using this and equation \eqref{eqn:ParCoeff} we find
\begin{align}
\|G^{\Lambda^{(2)}_{n+1}\backslash\Lambda^{(2)}_{n+1-\ell}}E_n\|^2 = &
	\frac{\tilde{\lambda}_2^{2n}\lambda_1^{2r(n)}\cdot
	\sum_{x_2=0}^{n-\ell_2} \tilde{\lambda}_2^{2x_2}
	\lambda_1^{2r(x_2)}}{\sum_{x_2 = 0}^{n-1}
	\tilde{\lambda}_2^{2x_2}\lambda_1^{2r(x_2)}\cdot \sum_{x_2
	= n-\ell_2+1}^{n-1} \tilde{\lambda}_2^{2x_2}\lambda_1^{2r(x_2)}} \nonumber \\
\leq & \label{GeoBound}
	\max(\lambda_1^4, \, \lambda_1^{-4}) \cdot
	\frac{\tilde{\lambda}_2^{2(\ell_2-1)}(1-\tilde{\lambda}_2^{2(n+1-\ell_2)})(1-\tilde{\lambda}_2^2)}{
	(1-\tilde{\lambda}_2^{2\ell_2})(1-\tilde{\lambda}_2^{2n})}.
\end{align}
The inequality follows from applying the bound $\min(1, \lambda_1^2)\leq
\lambda_1^{2r(x_2)} \leq \max(1, \lambda_1^2)$, and noticing \newline
$\frac{\max(1, \lambda_1^2)}{\min(1, \lambda_1^2)} = \max(\lambda_1^2,
\lambda_1^{-2})$. The upper bound in \eqref{GeoBound} is of the form described
in \eqref{CommonBound}. Therefore we define
\begin{equation*}
\epsilon_{\ell_2} = \max(\lambda_1^2, \, \lambda_1^{-2}) \min(1, \,
\tilde{\lambda}_2^{\ell_2-1}) \sqrt{\frac{1-\tilde{\lambda}_2^2
}{1-\tilde{\lambda}_2^{2\ell_2}}}.
\end{equation*}
for the smallest value of $\ell_2$ such that
$\epsilon_{\ell_2}^2<\frac{1}{\ell_2}$. All of the conditions for the
martingale method hold for this value of $\ell_2$, so the spectral gap
of $H^{\Lambda_L} = H^{\Lambda_N^{(2)}}$ is bounded below by
\begin{equation}
\gamma(\Lambda_L) \geq  \min_{0\leq b_2 \leq 2L-\ell_2}
\gamma\left(P(-\frac{m_2}{m_1}b_2,b_2,2L,\ell_2)  \right) \cdot
\frac{(1-\epsilon_{\ell_2}\sqrt{\ell_2})^2}{\ell_2}.
\end{equation}

\vskip 12pt
\textbf{Martingale Method in $x_1$ Direction:} Fix $b_2$ and $L$. We apply the
martingale method in the direction of $x_1$ to
$P(-\frac{m_2}{m_1}b_2,b_2,2L,\ell_2)$ to obtain a lower bound on
$\gamma_{\ell_2}$. Let $N = 2L$ and define $\Lambda_n^{(1)} :=
P(-\frac{m_2}{m_1}b_2, b_2, n, \ell_2).$ For these volumes to be connected, we
require $n \geq \frac{m_2}{m_1}+1$. For the same reason $\frac{m_2}{m_1}+1$ also
serves as a lower bound for $\ell_1$.

Condition (ii) is satisfied with $\gamma_{\ell_1}$ defined by
\begin{align*}
	\gamma_{\ell_1} &:= \min_{\ell_1 \leq n\leq N}
	\gamma(\Lambda_n^{(1)}\backslash\Lambda_{n-\ell_1}^{(1)})\\
	& =  \min_{-\frac{m_2}{m_1}b_2 \leq b_1 \leq L-\ell_1 - \frac{m_2}{m_1}b_2} \gamma\left(P(b_1,b_2,\ell_1,\ell_2)  \right)
\end{align*}
For any choice of $(b_1, b_2)$, the area of the parallelogram $P(b_1, b_2,
\ell_1, \ell_2)$ is $\ell_1 \ell_2$. If the $\ell_i$ are fixed, and
$\ell_1 \geq \tfrac{m_2}{m_1} + 1$, then there are only a finite number of
distinct connected subsets of $\bZ^2$ (up to translations) that are contained
in a parallelogram of area $\ell_1\ell_2$. Let
$\mathcal{P}(\vec{\ell})$ denote the collection of all such distinct subsets.
Since the PVBS models are translation invariant, it follows that
\begin{equation}
\gamma_{\ell_1}\geq \gamma(\vec{\ell}) \coloneqq \min_{P \in
\mathcal{P}(\vec{\ell})} \gamma(P).
\end{equation}
By definition, $\gamma(\vec{\ell})$ is nonzero and independent of $b$ and
$L$. Therefore, we can use $\gamma(\vec{\ell})$ for Condition (ii) in place of
$\gamma_{\ell_1}$.

For Condition (iii) recall that, by assumption, $\lambda_1 \neq
1$. Using similar calculations and bounds of the operator norm as computed in
the $x_2$ direction, we find
\begin{align*}
	\|G^{\Lambda^{(1)}_{n+1}\backslash\Lambda^{(1)}_{n+1-\ell}}E_n\|^2 
	= &
	\frac{\lambda_1^{2n}\sum_{x_1=0}^{n-\ell_1}\lambda_1^{2x_1}}
	{\sum_{x_1 = 0}^{n-1}\lambda_1^{2x_1}\cdot \sum_{x_1
= n-\ell_1+1}^{n-1} \lambda_1^{2x_2}} \\
	\leq & \min(1,
\, \lambda_1^{2(\ell_1-1)})\frac{1-\lambda_1^2}{1-\lambda_1^{2\ell_1}},
\end{align*}
Which is also of the form of \eqref{CommonBound}. Therefore, define
\begin{equation}
\epsilon_{\ell_1} := \min(1, \,
\lambda_1^{\ell_1-1})\sqrt{\frac{1-\lambda_1^2}{1-\lambda_1^{2\ell_1}}}
\end{equation}
for the smallest value of $\ell_1\geq \frac{m_2}{m_1} + 1$ such that
$\epsilon_{\ell_1}^2<\frac{1}{\ell_1}$. Then all conditions for the martingale
method are satisfied for this value of $\ell_1$ and $\gamma(\Lambda_N^{(1)})$
satisfies the nonzero lower bound:
\begin{equation}
\gamma(\Lambda_N^{(1)})\geq
\min_{P \in \mathcal{P}(\vec{\ell})}\gamma(P)\frac{(1-\epsilon_{\ell_1}\sqrt{\ell_1})^2}{\ell_1}.
\end{equation}

Since this result holds for all choices of $b_2$ and $L$, it follows
that this is also a lower bound for $\gamma_{\ell_2}$, which produces a uniform
lower bound on $\gamma(\Lambda_L)$. Applying Theorem \ref{thm:GNSSpecGap} on
$\gamma(D)$, we find
\begin{equation*}
	\gamma(D) \geq \min_{P \in \mathcal{P}(\vec{\ell})}\gamma(P)
	\frac{\left(1-\epsilon_{\ell_1}\sqrt{\ell_1}  \right)^2}{\ell_1} \cdot
	\frac{\left(1-\epsilon_{\ell_2}\sqrt{\ell_2}  \right)^2}{\ell_2} > 0.
\end{equation*}

\vskip12pt
\noindent
\textbf{Case 1b}: 
We still consider $(\log \lambda_1, \, \log\lambda_2)
\neq \|\log\vec{\lambda}\|(m_1, \, m_2)$, but now we assume that for all $j$
either $\lambda_j = 1$ or $m_j = 0$. Since we are also assuming that $(\log \lambda_1, \, \log\lambda_2) \neq (1,\, 1)$, as this is a gapless case, and $(m_1, \, m_2)
\neq (0, \, 0)$, as $m$ is a unit vector, this falls into two cases:
\[\lambda_1 \neq 1, \quad \lambda_2 = 1, \quad m_1 = 0, \quad m_2 = 1 \qquad
\text{or} \qquad \lambda_1 = 1, \quad \lambda_2 \neq 1, \quad m_1 = 1, \quad m_2
= 0 .
\]
We motivate the need to consider these cases separately. In the former case, the
definition for the slant boundaries of the parallelograms $P(b_1, \, b_2, \,
L_1, \, L_2)$ are undefined.
In the
latter case, the parallelograms $P(b_1, \, b_2, \, L_1, \, L_2)$ are rectangles.
Since in this case, $(\log\lambda_1, \, \log\lambda_2) = c(0, 1)$ is the outward
pointing normal to some boundary of $P(b_1, \, b_2, \, L_1, \, L_2)$, this is
not a good sequence of volumes to apply the martingale method.
\vskip12pt
\textbf{Volumes:} By Remark \ref{rem:coordinates}, it is sufficient to only
consider the case $\lambda_1 \neq 1, \, \lambda_2 = 1, \, m_1 = 0, \, m_2 = 1$. For this choice of $m$,
the infinite volume $D$ is the upper half plane $\{x\in\bZ^2 \, :\, x_2\geq 0\}$. Since $(\log\lambda_1, \, \log\lambda_2) = c(1, 0)$, we must
choose finite volumes with no vertical boundaries. So, we consider
parallelograms with the vector $v = (1,1)$ generating the slant
boundaries. Specifically,
\begin{align*}
	P(b_1,b_2,L_1,L_2) =\{x\in\mathbb{Z}^2: 0\leq v \cdot (x-b) < L_1,\ 0 \leq x_2 -b_2 < L_2 \}
\end{align*}
We choose $\Lambda_L = P(-L, \, 0, \, 2L, \, 2L)$ to be the sequence of volumes
increasing to $D$.  This is the simplest sequence of parallelograms to choose as
given $b_1, \, b_2, \, L_1,\, L_2$ are integer valued, the smallest and largest
values of $x_1$ value for any $x_2$ will lie on the boundary. Hence,
there are no remainders $r(x_2)$. As such, the proof for this case is the
similar to Case 1a, with several simplifications by our choice of $v$.

\vskip12pt
\textbf{Normalization Coefficient:} Since we no longer have a remainder $r(x_2)$
and since $\lambda_2 = 1$, the normalization constant calculated in
\eqref{eqn:ParCoeff} can be simplified to
 \begin{align*}
 	C(P(b_1,b_2,L_1,L_2)) 
	= \lambda^{2b} \sum_{x_2=0}^{L_2 -1}(\lambda_1^{-1})^{2x_2}\sum_{x_1=0}^{L_1 -1} \lambda_1^{2x_1}.
 \end{align*}
\vskip 12pt 
\textbf{Martingale Method in $x_2$ Direction:} Define $N = 2L$ and let
$\Lambda_n^{(2)} = P(-L, \, 0, \, 2L, \, n)$. Then $\Lambda_L =
\Lambda_N^{(2)}$. For any value of $\ell_2$, our choice of $v$ is such
that the volumes $\Lambda_{n}\backslash\Lambda_{n-\ell_2}$ are isomorphic
sublattices of $\bZ^2$. By the translation invariance of the PVBS models,
Condition (ii) of the martingale method is satisfied with 
\[\gamma_{\ell_2} = \min_{\ell_2 \leq n \leq N}
\gamma(\Lambda_n^{(2)}\backslash\Lambda_{n-\ell_2}^{(2)}) =
\gamma(\Lambda_{\ell_2}^{(2)}).\]

It only remains to check Condition (iii). Using Lemma
\ref{lem:EpsilonCalculation}, summing the geometric series and simplifying the
expression yields
\begin{equation*}\label{SpecialParCase}
\|G^{\Lambda_{n+1}^{(2)}\backslash \Lambda_{n+1-\ell_2}^{(2)}}\|^2 =
\frac{(1-\lambda_1^{-2(n+1-\ell_2)})(1-\lambda_1^{-2})
\lambda_1^{-2(\ell_2-1)}}{ (1-\lambda_1^{-2n})(1-\lambda_1^{-2\ell_2})},
\end{equation*}
which is of the form \eqref{CommonBound}. Therefore, we define
\begin{equation*}
\epsilon_{\ell_2} = \min(1, \, \lambda_1^{-(\ell_2-1)})
\sqrt{\frac{1-\lambda_1^{-2}}{1-\lambda_1^{-2\ell_2}}}.
\end{equation*}
It can be shown that the smallest value of $\ell_2$ such that
$\epsilon_{\ell_2}^2<\frac{1}{\ell_2}$ is $\ell_2=2$. Choosing this value of
$\ell_2$ gives
\begin{equation*}
\epsilon_2 = \frac{\min(1,\, \lambda_1^{-1})}{\sqrt{1+\lambda_1^{-2}}}.
\end{equation*}
Thus, we have the following lower bound on $\gamma(\Lambda_L):$
\begin{equation}\label{FirstBound}
\gamma(\Lambda_L)\geq
\gamma(\Lambda_{2}^{(2)})\cdot\frac{(1-\epsilon_2\sqrt{2})^2}{2},
\end{equation}
where $\Lambda_{2}^{(2)} = P(-L, 0, 2L, 2)$. 

\textbf{Martingale Method in $x_1$ Direction:} We now apply the martingale
method to $\Lambda_{2}^{(2)}$ to obtain a lower bound on
$\gamma(\Lambda_{2}^{(2)})$ that is independent of $L$. Let $N = 2L$ and
define $\Lambda_{n}^{(1)} = P(-L, 0, n, 2)$. Similar to the appliation in the
$x_2$ direction, Condition (ii) of the martingale method is satisfied for
\[
\gamma_{\ell_1} = \gamma(\Lambda_{\ell_1}^{(1)})\cong\gamma(P(0,0,\ell_1,2)).
\]

For Condition (iii), a similar calculation as the application in the direction
of $x_2$ shows
\begin{equation}
\|G^{\Lambda_{n+1}^{(1)}\backslash \Lambda_{n+1-\ell_1}^{(1)}}E_n\|^2 \leq
\frac{(1-\lambda_1^2)\min(1, \,
\lambda_1^{2(\ell_1-1)})}{1-\lambda_1^{2\ell_1}}.
\end{equation}
This is the same bound we had for the first application with
$\lambda_1^{-1}$ replaced with $\lambda_1$. Therefore, Condition (iii) once
again holds for $\ell_1 = 2$ and 
\[\epsilon_1 = \frac{\min(1,\,
\lambda_1)}{\sqrt{1+\lambda_1^{2}}}.\] Thus, we see that
\begin{equation}\label{2ndBound}
\gamma(\Lambda_{2}^{(2)}) \geq
\gamma(P(0,0,2,2))\frac{(1-\epsilon_1\sqrt{2})^2}{2}.
\end{equation}
Substituting this lower bound into \eqref{FirstBound} produces a lower bound on
$\gamma(\Lambda_L)$ that is independent of $L$. 

Therefore, we obtain
\begin{equation}
	\gamma(D) \geq \gamma\left(P(0,0,2,2)\right)
	\frac{\left(1-\epsilon_1\sqrt{2}\right)^2}{2} \cdot 
	\frac{\left(1-\epsilon_2\sqrt{2}\right)^2}{2} > 0.
\end{equation}

\vskip12pt
\noindent
\textbf{Case 2a}: We now consider $(\log\lambda_1,\log\lambda_2) =
\|\log\vec{\lambda}\|(m_1,m_2)$. We invoke the reflection symmetry of the model
to produce a boundary for which $m_1\geq 0$ and $m_2 \geq 0$, see Remark
\ref{rem:coordinates}.
For this case, we also
assume that $m_1 \neq 0$ and $m_2\neq 0$. Since $\log\vec{\lambda} =
\|\log\vec{\lambda}\|\cdot m$, this implies that $\lambda_1 > 1$ and $\lambda_2
>1$.

\vskip12pt
\textbf{Volumes:} We cannot choose parallelograms for the sequence 
$\Lambda_L$, since $(\log\lambda_1,\log\lambda_2) = c(m_1,m_2)$ would be
an outward pointing normal to the boundary opposite to $D$. 
We instead use trapezoids which have one boundary along the boundary of $D$, and
vertical or horizontal lines for the other boundaries. The general
trapezoid of interest is defined by
\begin{align*}
	T(b_1,b_2,L_1,L_2) =\{x\in\mathbb{Z}^2: 0\leq \frac{m}{m_1} \cdot (x-b),\ x_1-b_1 < L_1,\ 0 \leq x_2 -b_2 < L_2 \}
\end{align*}
\begin{figure}[h]
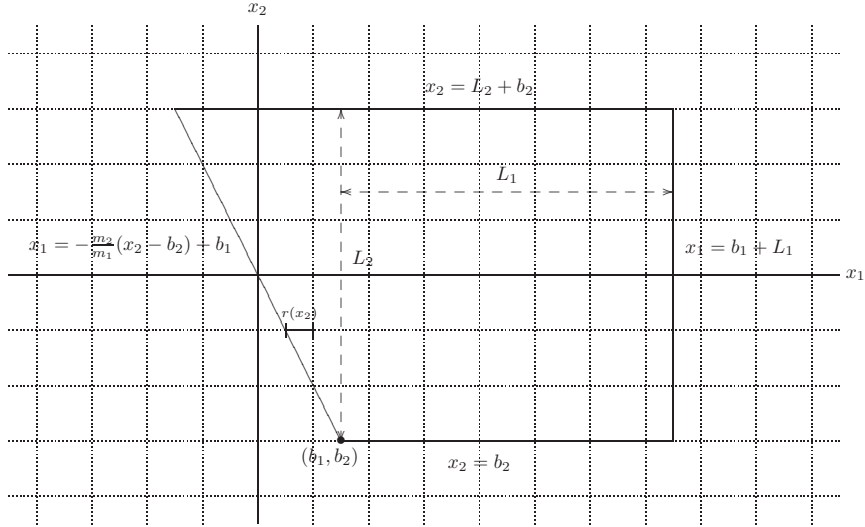

\begin{adjustbox}{max totalsize={.7\textwidth}{.7\textheight}}
{
\xy
(0,-25); (0,65) ** \dir{-}; 
(0,68)*+{x_2};
(-45,20); (105,20) ** \dir{-}; 
(108,20)*+{x_1};
(-40,-25);(-40,65) **\dir{.}; 
(-30,-25);(-30,65) **\dir{.};
(-20,-25);(-20,65) **\dir{.};
(-10,-25);(-10,65) **\dir{.};
(10,-25);(10,65) **\dir{.};
(20,-25);(20,65) **\dir{.};
(30,-25);(30,65) **\dir{.};
(40,-25);(40,65) **\dir{.};
(50,-25);(50,65) **\dir{.};
(60,-25);(60,65) **\dir{.};
(70,-25);(70,65) **\dir{.};
(80,-25);(80,65) **\dir{.};
(90,-25);(90,65) **\dir{.};
(100,-25);(100,65) **\dir{.};
(-45,-20); (105,-20) ** \dir{.}; 
(-45,-10); (105,-10) ** \dir{.}; 
(-45,10); (105,10) ** \dir{.};
(-45,0); (105,0) ** \dir{.};
(-45,30); (105,30) ** \dir{.};
(-45,40); (105,40) ** \dir{.};
(-45,50); (105,50) ** \dir{.};
(-45,60); (105,60) ** \dir{.};
(-15,50); (15,-10)**\dir{-}; 
(75,50); (75,-10)**\dir{-};
(-15,50); (75,50)**\dir{-};
(15,-10); (75,-10)**\dir{-};
(87,25)*+{x_1=b_1+L_1};
(-23,25)*+{x_1=-\frac{m_2}{m_1}(x_2-b_2)+b_1};
(40,54)*+{x_2 = L_2+b_2};
(40,-14)*+{x_2 = b_2};
(15,-10)*+{\bullet};
(13,-13)*+{(b_1, b_2)};
(5,10); (10,10)**\dir{-};
(5,8.75); (5,11.25)**\dir{-};
(10,8.75); (10,11.25)**\dir{-};
(7.5,13)*+{\scriptstyle r(x_2)}; 
{\ar@{<-->} (15,-10)*{}; (15,50)*{}};
(19,23)*+{L_2};
{\ar@{<-->} (15,35)*{}; (75,35)*{}};
(45,38)*+{L_1};

\endxy 
}
\end{adjustbox}
   \caption{A general trapezoid $T(b_1, b_2, L_1, L_2)$.}
  \label{fig:trapezoid}
\end{figure}
where $b =(b_1,b_2)$ is the base point of the trapezoid and $L_1,\ L_2$
are the lengths in the coordinate directions, see Figure \ref{fig:trapezoid}. We
always choose $b_2, \, L_1, \, $ and $L_2$ to be integer valued. However, $b_1$ will typically not be
integer valued. For the sequence $\Lambda_L\nearrow D$, we choose
$\Lambda_L = T(\frac{m_2}{m_1}L, -L, 2L, 2L).$

\vskip12pt
\textbf{Normalization Coefficient:} For a fixed integer value of $x_2$, the
smallest integer value $x_1^{min}$ such that $(x_1^{min}, \, x_2)$ is contained
in the trapezoid $T(b_1,b_2,L_1,L_2)$ is given by
\begin{equation*}
x_1^{min} = -\frac{m_2}{m_1}(x_2-b_2)+b_1+r(x_2)
\end{equation*}
where $r(x_2)\in[0,1)$. In particular, since  $b_1$ need not be integer valued,
it follows that $x_1^{min}(b_2)=b_1+r(b_2)$. For all $x_2$, the largest integer
value of $x_1$, denoted $x_1^{max}$, contained in the trapezoid $T(b_1, \, b_2,
\, L_1, \, L_2)$ is
\begin{equation*}
x_1^{max} = L_1 + b_1 - 1 + r(b_2). 
\end{equation*}
Since $\log \lambda_1 = cm_1 > 1$ and $\log\lambda_2 = cm_2$ with $c>0$, it
follows that $\lambda_2\cdot\lambda_1^{-m_2/m_1} = 1$ and we find
\begin{align}
C(T(b_1, \, b_2, \, L_1, \, L_2)) 
	& = \sum_{x_2=b_2}^{b_2+L_2-1} \lambda_2^{2x_2} \sum_{x_1 = b_1
		-\tfrac{m_2}{m_1}(x_2-b_2)+r(x_2)}^{L_1+b_1 -1 +r(b_2)}	\lambda_1^{2x_1}
		\nonumber\\
	& = \lambda^{2b} \sum_{x_2 = 0}^{L_2 - 1}
	\lambda_1^{2 r(x_2+b_2)}\, \cdot \, 
	\frac{\lambda_1^{2(L_1+r(b_2)+\frac{m_2}{m_1}x_2 -r(x_2+b_2))} - 1}{\lambda_1^2.
	- 1} \label{NormCoeffTrap}
\end{align}
Define
\[C_{max} = \frac{1}{\lambda_1^2-1} \quad \text{and} \quad C_{min} =
\frac{1-\lambda_1^{-2}}{\lambda_1^2-1}.\]
We show that
\begin{equation}\label{BoundsOnFrac}
C_{min}\lambda_1^{2(L_1+r(b_2)+\frac{m_2}{m_1}x_2 -r(x_2+b_2))} 
	\leq
\frac{\lambda_1^{2(L_1+r(b_2)+\frac{m_2}{m_1}x_2 -r(x_2+b_2))} - 1}{\lambda_1^2- 1}
	\leq
C_{max}\lambda_1^{2(L_1+r(b_2)+\frac{m_2}{m_1}x_2 -r(x_2+b_2))} 
\end{equation}
The upper bound follows immediately. For the lower bound, note that
\begin{equation*}
L_1 + r(b_2) + \frac{m_2}{m_1}x_2 - r(x_2+b_2) = L_1 + x_1^{min}(b_2) -
x_1^{min},
\end{equation*}
where $x_1^{min}$ is the minimum $x_1$ value for $x_2+b_2$. Since $m_1, \, m_2
>0$, as $x_2$ increases the minimum value of $x_1$
decreases. Therefore, $x_1^{min}(b_2)-x_1^{min}>0$. Trivially $L_1\geq 1$, and
so
\begin{equation*}
1 \leq L_1 + r(b_2) +\frac{m_2}{m_1}x_2 - r(x_2+b_2)
\end{equation*}
and the lower bound follows. Inserting these bounds into \eqref{NormCoeffTrap}
yields
\begin{equation}\label{UpLowBoundTrapCoeff}
C_{min}\lambda^{2b} \lambda_1^{2(L_1+r(b_2))}
\sum_{x_2=0}^{L_2-1}\lambda_2^{2x_2}
	\leq
C(T(b_1,b_2,L_1,L_2)
	\leq
 C_{max} \lambda^{2b}\lambda_1^{2(L_1+r(b_2))}
\sum_{x_2=0}^{L_2-1}\lambda_2^{2x_2}
\end{equation}
\vskip12pt
\textbf{Martingale Method in $x_2$ Direction:} Let $N=2L$ and set
$\Lambda_n^{(2)} = T(\frac{m_2}{m_1}L, -L, 2L, n).$ Then for $n \geq
\ell_2$,

\begin{equation*}
\Lambda_n^{(2)}\backslash \Lambda_{n-\ell_2}^{(2)} =
T(-\frac{m_2}{m_1}(n-L-\ell_2), \, n-L-\ell_2, \, 2L+\frac{m_2}{m_1}(n-\ell_2),
\, \ell_2 ).
\end{equation*}
Therefore, Condition (ii) of the martingale method is satisfied for
\begin{align*}
	\gamma_{\ell_2} &:= \min_{\ell_2\leq n\leq N}
	\gamma\left(T(-\frac{m_2}{m_1}(n-L-\ell_2), \, n-L-\ell_2, \,
	2L+\frac{m_2}{m_1}(n-\ell_2), \, \ell_2 )\right)\\
	& =  \min_{-L\leq b_2 \leq 2L-\ell_2}
	\gamma\left(T(-\frac{m_2}{m_1}b_2, \, b_2, \, 2L+\frac{m_2}{m_1}(L+b_2),
	\,\ell_2) \right)
\end{align*}
For Condition (iii), notice that for all $k<n$
\begin{equation*}
\Lambda_n^{(2)} \backslash \Lambda_k^{(2)} = T(-\frac{m_2}{m_1}(k-L), k-L,
2L+\frac{m_2}{m_1}k, n-k).
\end{equation*}
Using \eqref{UpLowBoundTrapCoeff} for each of the four normalization constants
we compute
\begin{equation}
\|G^{\Lambda^{(2)}_{n+1}\backslash\Lambda^{(2)}_{n+1-\ell}}E_n\|^2 \leq 
	\frac{C_{max}^2}{C_{min}^2}\cdot
	\frac{\lambda_2^{2(\ell_2-1)}(1-\lambda_2^2)(1-\lambda_2^{2(n+1-\ell_2)})}{(1-\lambda_2^{2\ell_2})(1-\lambda_2^{2n})}
	\label{ExactTrap}
\end{equation}
This is of the same form as \eqref{CommonBound}. Since $\lambda_2>1$, we choose
\begin{equation}
\epsilon_{\ell_2} := \frac{C_{max}}{C_{min}}
	\sqrt{\frac{1-\lambda_2^2}{1-\lambda_2^{2\ell_2}}},
\end{equation}
for the smallest value of $\ell_2$ satisfying
$\epsilon_{\ell_2}^2<\frac{1}{\ell_2}$. Therefore, Condition (iii) of the
martingale method is also satisfied and
\begin{equation}
\gamma(\Lambda_L) \geq \min_{-L\leq b_2 \leq 2L-\ell_2}
	\gamma\left(T(-\frac{m_2}{m_1}b_2, \, b_2, \, 2L+\frac{m_2}{m_1}(L+b_2),
	\,\ell_2) \right)\cdot 
\frac{(1-\epsilon_{\ell_2}\sqrt{\ell_2})^2}{\ell_2}.
\end{equation}
\vskip12pt
\textbf{Martingale Method in $x_1$ Direction:} We apply the martingale method
to produce a lower bound on $\gamma_{\ell_2}$. Fix $L$ and $b_2$, let $N = 2L +
\tfrac{m_2}{m_1}(L+b_2)$, and define $\Lambda_{n}^{(1)} = T(-\frac{m_2}{m_1}b_2,
b_2, n, \ell_2).$ 

For Condition (ii), recall that the martingale method assumes
that $\Lambda_0^{(1)} = \emptyset$. So, for any $\ell_1\geq 2$, 
\[\Lambda_{\ell_1}\backslash \Lambda_0 = T(-\frac{m_2}{m_1}b_2, b_2, \ell_1,
\ell_2).\]
Let
\begin{equation}\label{GammaT}
\gamma_T(\vec{\ell}) = \min_{b\in\bZ^2}\gamma(T(b_1, b_2,
\ell_1, \ell_2)).
\end{equation} 
Similar to the parallelogram case, there are only a finite number of distinct
(up to translations) trapezoids $T(b_1, b_2, \ell_1, \ell_2)$ contained in
$\bZ^2$. Therefore, by the translation invariance of the PVBS model, the minimum
in \eqref{GammaT} is positive, and
$\gamma(\Lambda_{\ell_1}\backslash\Lambda_0)\geq
\gamma_T(\vec{\ell})>0$. Furthermore, for $n>1$
\begin{equation*}
\Lambda_{n+\ell_1}\backslash\Lambda_n = B(-\frac{m_2}{m_1}b_2+n-\ell_1, b_2,
\ell_1, \ell_2) 
\end{equation*}
where $B(b_1,b_2, \ell_1, \ell_2 )$ is the rectangular box
\[B(b_1, b_2, \ell_1, \ell_2) = \{x\in\bbZ^2 : b_1 \leq x < b_1+\ell_1,
\, b_2\leq x_2< b_2+\ell_2\}.
\]
Regardless of the values of $b_1$ and $b_2$, $B(b_1, b_2,
\ell_1, \ell_2) \cong B(0, 0, \ell_1, \ell_2)$. Therefore,
$\gamma(\Lambda_{n+\ell_1}\backslash\Lambda_n)~=~\gamma(B(0,0,\ell_1,\ell_2))$
and Condition (ii) of the martingale method is satsified for
\begin{equation}
\gamma_{\ell_1} = \min\left(\gamma_T(\vec{\ell}), \, \gamma
(B(0,0,\ell_1, \ell_2))\right).
\end{equation}

For Condition (iii), since for all $n>k$, $\Lambda_n\backslash\Lambda_k$ is a
rectangular box
\begin{equation}
\frac{C(\Lambda_{n+1}^{(1)}\backslash\Lambda_{n}^{(1)})}{C(\Lambda_{n+1}^{(1)}
\backslash\Lambda_{n+1-\ell_1}^{(1)})} =
\frac{\lambda_1^{2(\ell_1-1)}(1-\lambda_1^2)}{1-\lambda_1^{2\ell_1}}.
\end{equation}
Using \eqref{UpLowBoundTrapCoeff}, we additionally obtain the bound
\begin{equation}
\frac{C(\Lambda_{n+1-\ell_1}^{(1)})}{C(\Lambda_{n}^{(1)})} \leq
\frac{C_{max}}{C_{min}}\lambda_1^{-2(\ell_1-1)}.
\end{equation}
Putting these together yields
\begin{equation}
\|G^{\Lambda_{n+1}^{(1)}\backslash\Lambda_{n+1-\ell_1}^{(1)}}E_n\|^2 \leq
\frac{C_{max}}{C_{min}}\cdot\frac{1-\lambda_1^2}{1-\lambda_1^{2\ell_1}} :=
\epsilon_{\ell_1}^2.
\end{equation}
which decays exponentially. 

Choosing $\ell_1$ to be the smallest integer such
that $\epsilon_{\ell_1}^2< \frac{1}{\ell_1}$ satisfies all the conditions for the martingale method and
\begin{equation}
\gamma\left(T(-\frac{m_2}{m_1}b_2, \, b_2, \, 2L+\frac{m_2}{m_1}(L+b_2),
\,\ell_2)\right) \geq \min\left(\gamma_T(\vec{\ell}), \, \gamma(B(0,0,\ell_1,
\ell_2))\right) \cdot \frac{(1-\epsilon_{\ell_1}\sqrt{\ell_1})^2}{\ell_1} > 0.
\end{equation}
Since this bound is independent of the choice $L$ and $b_2$, replacing this for
$\gamma_{\ell_2}$ in the lower bound for $\gamma(\Lambda_L)$ and applying
Theorem proves
\ref{thm:GNSSpecGap}
\begin{equation*}
\gamma(D) \geq \min\left(\gamma_T(\vec{\ell}), \, \gamma (B(0,0,\ell_1,
\ell_2))\right) \frac{(1-\epsilon_{\ell_2}\sqrt{\ell_2})^2}{\ell_2} \cdot
\frac{(1-\epsilon_{\ell_1}\sqrt{\ell_1})^2}{\ell_1} > 0.
\end{equation*}

\vskip12pt
\noindent
\textbf{Case 2b:} We now consider $(\log\lambda_1,\log\lambda_2) =
\|\log\vec{\lambda}\|(m_1,m_2)$ and either $m = (1,0)$ or $m = (0,1)$. These
cases are considered separately from the those covered in Case 2a for the
following reasons. In the case that $m = (0,1)$, the trapezoids $T(b_1, b_2, L_1, L_2)$
are undefined. If $m = (1,0)$, the boundary of $D$ becomes a vertical line, and
the trapezoid $T(b_1, b_2, L_1, L_2)$ is the rectangular box $B(b_1, b_2, L_1,
L_2)$. Since $\log \vec{\lambda} = \|\log\vec{\lambda}\|m$ is an outward
pointing normal to one of the boundaries of $B(b_1, b_2, L_1, L_2)$, this is not
a good choice for the volume sequence for the PVBS model.

Permuting the coordinate indices, we need only prove this special case $m
= (1, 0)$. This implies $\lambda_1 > 1$ and $\lambda_2 = 1$. 

\vskip12pt
\textbf{Volumes:} We modify the trapezoid $T(b_1, b_2, L_1, L_2)$ from Case 2a
so that the right most boundary is slanted with slope determined by the vector
$v = (1, -1)$. Specifically, the trapezoids are defined as
\begin{equation}
	T(b_1,b_2,L_1,L_2) =\{x\in\mathbb{Z}^2: 0\leq x_1-b_1 ,\ v \cdot
	(x-b)< L_1,\ 0 \leq x_2 -b_2 < L_2 \}.
\end{equation}
Since the boundary of $D$ is defined by $x_1 = 0$, we choose as our sequence of
increasing volumes $\Lambda_L = T(0,-L,2L,2L).$ In this case, we only
consider trapezoids for which $b_1$, $b_2$, $L_1$, and $L_2$ are integer
valued.

\vskip12pt
\textbf{Normalization Coefficient:}
We bound the normalization constant for the one particle ground state on
$T(b_1,b_2,L_1,L_2)$. Since all $b_1$, $b_2$, $L_1$, and $L_2$ are integer
valued, there are no remainders. For a fixed value of $x_2$, the smallest value
of $x_1$ is always $x_{1}^{min} = b_1$, and the largest value of $x_1$
is $x_1^{max} = b_1+ L_1 + x_2 - b_2 - 1$. Using $\lambda_2 =1$, we obtain
\begin{align*}
C(T(b_1, b_2, L_1, L_2)) 
& = \lambda^{2b}\sum_{x_2 = 0}^{L_2-1} \frac{1 -
	\lambda_1^{2(L_1+x_2)}}{1-\lambda_1^2}.
\end{align*}
Since $\lambda_1 > 1$, similar to Case 2a, we have
\begin{equation}\label{UpLowBoundsTrap2}
C_{min}\lambda^{2b} \lambda_1^{2L_1}\sum_{x_2 = 0}^{L_2-1}
\lambda_1^{2x_2} \leq C(T(b_1, b_2, L_1, L_2)) \leq
C_{max} \lambda^{2b}  \lambda_1^{2L_1}\sum_{x_2 = 0}^{L_2-1}
\lambda_1^{2x_2},
\end{equation}
where $C_{min}$ and $C_{max}$ are defined as they were in Case 2a.

\vskip12pt
\textbf{Martingale Method in $x_2$ Direction:}
Let $N = 2L$ and set $\Lambda_n^{(2)} = T(0, -L, 2L, n).$ Then
$\Lambda_n^{(2)}\nearrow \Lambda_N^{(2)}$. For $k<n$,
\begin{equation}\label{SetDiff2b}
\Lambda_{n}^{(2)} \backslash \Lambda_k^{(2)} = T(0, k-L, k+2L, n-k).
\end{equation}
Therefore, Condition (ii) is satisfied with
\begin{align*}
\gamma_{\ell_2} & = \min_{\ell_2 \leq n \leq 2L} \gamma(\Lambda_{n}^{(2)}
\backslash \Lambda_{n-\ell_2}^{(2)}) \\\
	& = \min_{\ell_2 \leq n \leq 2L} \gamma(T(0, n-\ell_2-L, n-\ell_2+2L, \ell_2))\\
	& = \min_{0 \leq b_2 \leq 2L-\ell_2} \gamma(T(0, b_2-L, b_2 +2L, \ell_2))
\end{align*}

Using \eqref{UpLowBoundsTrap2}, we find
\begin{align*}
\frac{C(\Lambda_n^{(2)})}{C(\Lambda_{n+1-\ell_2}^{(2)})} & \leq
\frac{C_{max}}{C_{min}} \frac{\lambda_1^{2(n+1-\ell_2)}-1}{\lambda_1^{2n}-1} 
		\leq \frac{C_{max}}{C_{min}}\lambda_1^{-2(\ell_2-1)} 
		 = \frac{\lambda_1^4}{\lambda_1^2-1}  \lambda_1^{-2\ell_2}.	 
\end{align*}
Since the one particle ground state normalization coefficient increases as the
lattice increases we know that
\begin{equation}\label{BetterBound}
\frac{C(\Lambda_{n+1}^{(2)}\backslash\Lambda_n^{(2)})}{C(\Lambda_{n+1}^{(2)}
\backslash\Lambda_{n+1-\ell_2}^{(2)})} \leq 1.
\end{equation}
Therefore,
\begin{equation}
\|G^{\Lambda_{n+1}^{(2)}\backslash \Lambda_{n+1-\ell_2}^{(2)}}E_n\|^2 \leq
\frac{\lambda_1^4}{\lambda_1^2-1} \lambda_1^{-2\ell_2} \coloneqq
\epsilon_{\ell_2}^2,
\end{equation}
which decays exponentially. Letting $\ell_2$ be the smallest integer such that
$\epsilon_{\ell_2}^2 < \frac{1}{\ell_2},$ the martingale method yields
\begin{equation}
\gamma(\Lambda_L) \geq 	\min_{0 \leq b_2 \leq 2L-\ell_2} \gamma\left(T(0, b_2-L,
b_2 +2L, \ell_2)\right) \cdot 
\frac{(1-\epsilon_{\ell_2}\sqrt{\ell_2})^2}{\ell_2}.
\end{equation}

\vskip12pt
\textbf{Martingale Method in $x_1$ Direction:}
We apply the martingale method to the finite volume \newline $T(0, b_2-L,
b_2+2L, \ell_2)$ for a fixed $L$ and $b_2$ to obtain a lower bound on
$\gamma_{\ell_2}$. 

Let $N = b_2 + 2L$, and defined $\Lambda_n^{(1)} =
T(b_2+2L-n, b_2-L, n, \ell_2).$ Then $\Lambda_N^{(1)} = T(0, b_2-L, b_2+2L,
\ell_2)$ as desired. Similar to Case 2a, since $\Lambda_{0}^{(1)}$ is the empty
set, we have
\begin{align*}
\Lambda_{\ell_1}^{(1)}\backslash \Lambda_{0}^{(1)}  = T(b_2+2L-\ell_1, b_2-L,
\ell_1, \ell_2) 
& \cong T(0,0,\ell_1, \ell_2),
\end{align*}
where the last equivalence holds since $b_2$ and $L$ are integers. For $n\geq
1$
\begin{align*}
\Lambda_{n+\ell_1}^{(1)}\backslash \Lambda_{n}^{(1)} = B(b_2+2L-n, b_2-L,
\ell_1, \ell_2) 
& \cong B(0,0,\ell_1, \ell_2).
\end{align*}
Using the translation invariance of the PVBS model, Condition (ii) is satisfied
for
\begin{equation}
\gamma_{\ell_1} = \min\left( \gamma(T(0,0,\ell_1,\ell_2)),
\, \gamma(B(0,0,\ell_1,\ell_2)) \right).
\end{equation}

For Condition (iii), using $\lambda_2 = 1$ and $C(\Lambda_n\backslash\Lambda_m)
= C(\Lambda_n) - C(\Lambda_k)$ for $n\geq k \geq 1$, we can exactly calculate that
\begin{equation*}
C(\Lambda_n^{(1)}\backslash \Lambda_k^{(1)}) \quad = \quad \sum_{x_2 = b_2 -
L}^{\ell_2 +b_2 - L -1} \sum_{x_1 = b_2+2L-n}^{b_2+2L-k-1}\lambda_1^{2x_1} \quad
= \quad
\ell_2\lambda_1^{2(b_2+2L-n)}\frac{1-\lambda_1^{2(n-k)}}{1-\lambda_1^2}
\end{equation*}
It follows that,
\begin{equation*}
\frac{C(\Lambda_{n+1}^{(1)}\backslash
\Lambda_n^{(1)})}{C(\Lambda_{n+1}^{(1)}\backslash \Lambda_{n+1-\ell_1}^{(1)})} =
\frac{1-\lambda_1^2}{1-\lambda_1^{2\ell_1}}.
\end{equation*}
Since $\Lambda_{n+1-\ell_1}^{(1)} \subseteq \Lambda_{n}^{(1)}$, it follows that
$C(\Lambda_{n+1-\ell_1}^{(1)}) \backslash C(\Lambda_{n}^{(1)})\leq 1$, so
\begin{equation*}
\|G^{\Lambda_{n+1}^{(1)}\backslash \Lambda_{n+1-\ell_1}^{(1)}}E_n\|^2 \leq
\frac{1-\lambda_1^2}{1-\lambda_1^{2\ell_1}},
\end{equation*}
which decays exponentially. Let $\ell_1$ be the smallest integer such that
\begin{equation*}
\epsilon_{\ell_1}^2 \coloneqq \frac{1-\lambda_1^2}{1-\lambda_1^{2\ell_1}} <
\frac{1}{\ell_1}.
\end{equation*}

Therefore the three conditions of the martingale method are satisfied and
\begin{equation*}
\gamma(T(0, b_2-L, b_2+2L, \ell_2)) \geq 
	\min\left( \gamma(T(0,0,\ell_1,\ell_2)),\, \gamma(B(0,0,\ell_1,\ell_2)) \right)\cdot
	\frac{(1-\epsilon_{\ell_1}\sqrt{\ell_1})^2}{\ell_1}.
\end{equation*}
Since the choice of $\ell_1$ is independent of $L_2$ and $b_2$, substituting
this into the bound for $\gamma(\Lambda_L)$ gives
\begin{equation}
\gamma(\Lambda_L)\geq \min\Big( \gamma(T(0,0,\ell_1,\ell_2)),\,
\gamma(B(0,0,\ell_1,\ell_2))\Big)
\frac{(1-\epsilon_{\ell_1}\sqrt{\ell_1})^2}{\ell_1}\cdot \frac{(1-\epsilon_{\ell_2}\sqrt{\ell_2})^2}{\ell_2} > 0.
\end{equation}
Since $\Lambda_L \nearrow D$, the same bound holds for the spectral gap of
$H^D,$ as desired.


\section{Existence of a Spectral Gap in $d$ Dimensions}\label{sec:dDimCase}

In analogy with the analysis for $d=2$ in the previous section, we divide the
proofs for arbitrary $d$ into several cases. Cases 3a and 3b cover the situations
in which the vectors $m$ and $\log\vec\lambda$ are not parallel, that is
$\log\vec{\lambda} \neq \pm\Vert\log\vec{\lambda}\Vert m$.
For these cases, we choose volumes with $d$ pairs of parallel boundaries, i.e.
the $d$ dimensional analogue of parallelograms, sometimes called parallelotopes.
Case 3a is the case where there exists $j$ such that $m_j\neq 0$ and
$\lambda_j \neq 1$ and Case 3b is the case where there does not exist such
$j$. Case 4 deals with  $\log\vec{\lambda} = \Vert\log\vec{\lambda}\Vert m$. For
this case, we choose the $d$ dimensional analogue of the trapezoids in Case 2a
and Case 2b. We will refer back to arguments and calculations in the proofs for
two dimensions.

The proof of Corollary \ref{cor:GappedZd} is given at the end of the section.

\vskip12pt
\noindent
\textbf{Case 3a}: Suppose $\log\vec{\lambda} \neq \pm
\Vert\log\vec{\lambda}\Vert m$ and there exists $j\in\{1,\dots,d\}$ such that
$\lambda_j \neq 1$ and $m_j \neq 0$.  By Remark \ref{rem:coordinates}, we can
permute the indices so $\lambda_1 \neq 1$ and $m_1\neq 0$, without loss of
generality.

\vskip12pt
\textbf{Volumes:} As in Case 1a, we choose one pair of boundaries given by
$0\leq \frac{m}{m_1}\cdot x<L$. If we chose $-L\leq x_k <L$ for the other
boundaries, the volumes may not satisfy the heuristic that $\lambda^{x}$ is
maximized at a single point. This occurs exactly when
$(\log\lambda_1,\log\lambda_j) = c(m_1,m_j)$ as the ground state coefficients
$\lambda^x$ are maximized along some $(x_1,x_j)$ plane contained in the volume.
To avoid this issue, we introduce a different pair of slanted boundaries.
The non-parallel condition $\pm \Vert\log\vec{\lambda}\Vert m$ implies there
must be $\lambda_j$ such that $\lambda_1^{-m_j/m_1}\lambda_j \neq 1$.
The argument is as follows. 

The collection of vectors $\{ -m_j  e_j + m_1  e_1:
j=2, \dots, d\}$ is a linearly independent set such that all vectors are
orthogonal to $m$. Consequently, this set spans the perpendicular subspace of
$m$. If $\log\vec{\lambda}$ is not parallel to $m$, then it must have a non-zero
projection to this perpendicular space. Therefore, the inner product of
$\log\vec{\lambda}$ and $ -m_j  e_j + m_1  e_1$ is non-zero for some $j=2,
\ldots, d$. Then $\lambda_1^{-m_j/m_1}\lambda_j =
\exp[\frac{1}{m_1}(-m_j\log\lambda_1 + m_1\log\lambda_j)] \neq 1$. We permute
the indices so that this holds for $j=2$.

The extra pair of slanted boundaries is $-L \leq v\cdot x<L$ where the vector
$v=(0,1,v_3, \dots v_d)$ is defined as
\begin{displaymath}
   v_j := \left\{
     \begin{array}{lr}
       0 & \text{ if }\lambda_1^{-\frac{m_j}{m_1}}\lambda_j \neq 1\\
       -1 & \text{ if }\lambda_1^{-\frac{m_j}{m_1}}\lambda_j = 1\\
     \end{array}
   \right.
\end{displaymath} 
We apply the martingale method to the sequence of volumes is given by
\begin{align*}
	\Lambda_L:=\{x\in\mathbb{Z}^d:\ 0\leq \frac{m}{m_1}\cdot x <L,\ -L\leq v\cdot x
	<L,\ \text{and for } j\geq 3,\ -L\leq x_j< L\}.
\end{align*}

\textbf{Normalization Coefficient:}
To abbreviate notation, define 
\begin{align*}
	\tilde{\lambda}_1 :=\lambda_1,\quad
	\tilde{\lambda}_2 :=\lambda_1^{-\frac{m_2}{m_1}}\lambda_2,\quad\text{and}\quad
	\tilde{\lambda}_j :=(\lambda_1^{-\frac{m_j}{m_1}}\lambda_j) (\lambda_1^{-\frac{m_2}{m_1}}\lambda_2)^{-v_j} \text{ for } j\geq 3.
\end{align*}
By the choice of $v_j$, we guarantee that each $\tilde{\lambda}_j \neq 1$ for $j = 1, \dots, d$. 
We calculate the normalization constant for general volume 
\begin{align*}
	\Lambda':= \{x\in D:\ &b_1\leq \frac{m}{m_1}\cdot x <b_1+ L_1,\ b_2 \leq v\cdot x <b_2 + L_2,\ \text{for } j\geq 3, b_j\leq x_j  < b_j +  L_j \},
\end{align*}
which includes all possible subsets generated by the martingale method. For an
integer choice of $b_2$, the compound inequality $b_2 \leq v\cdot x <b_2 + L_2$
provide integer bounds on the $x_2$ coordinates of the form
\begin{align*}
	b_2+ \sum_{j=3}^d (-v_j)x_j\leq x_2< b_2+ L_2 -1 + \sum_{j=3}^d (-v_j)x_j.
\end{align*}
The compound inequality $b_1\leq \frac{m}{m_1}\cdot x <b_1+ L_1$ bounds the
$x_1$ coordinates. Equivalent integer bounds are given by adding a remainder
term $r$ such that $\sum_{j=2}^d -\frac{m_j}{m_1}x_j +  r$ equals the integer
$\lceil\sum_{j=2}^d -\frac{m_j}{m_1}x_j\rceil$.  Note that $r$ depends on $m$
and $x_2, \dots, x_d$, but not $x_1$.
Hence, the $x_1$ values are bounded by
\begin{align*}
	b_1 +  \sum_{j=2}^d -\frac{m_j}{m_1}x_j  + r \leq x_1 < b_1+L_1 + \sum_{j=2}^d
	\frac{-m_j}{m_1}x_j +r.
\end{align*}
To simplify notation in the calculation of the normalization coefficients, let $a_1(x)=\sum_{j=2}^d -\frac{m_j}{m_1}x_j$ and $a_2(x) =  \sum_{j=2}^d -\frac{m_j}{m_1}x_j + r$.  On can check that the normalization coefficient for $\Lambda'$ is then given by
\begin{align}
C(\Lambda') 	= \left(\prod_{j=1}^d \tilde\lambda_j^{2b_j}  \right)\sum_{x_d=0}^{L_d - 1} 
			\tilde{\lambda}_d^{2x_d}\sum_{x_{d-1}=0}^{L_{d-1} - 1}  \tilde{\lambda}_{d-1}^{2x_{d-1}} \dots\sum_{x_2=0}^{L_2- 1}  \tilde{\lambda}_2^{2x_2}\lambda_1^{2r}\sum_{x_1=0}^{L_1 -1} \tilde{\lambda}_1^{2x_1}.
\end{align}
This is bounded above and below by a constant multiple of a product of
geometric series after bounding the remainder term $\lambda_1^{2r}$
appropriately.

\textbf{Martingale Method in Each Direction:} 
We apply the martingale method $d$ times.  We first apply it to the sequence of volumes
\begin{align*}
	\Lambda^{(d)}_n := \Lambda_L \cap \{x_d<n-L\},
\end{align*}
for $\ell_d \leq n \leq 2L$ to generate a lower bound of the form
\begin{align*}
	\gamma(\Lambda^{(d)}_n) \geq \min_{-L+\ell_d \leq n \leq L}\gamma(\Lambda^{(d)}_n\setminus\Lambda^{(d)}_{n-\ell_d}) \frac{(1-\epsilon_d\sqrt{\ell_d})^2}{\ell_d}.
\end{align*}
Introducing $b_d = n-\ell_d$, we apply the martingale method to each $\Lambda^{(d)}_{b_d+\ell_d} \setminus\Lambda^{(d)}_{b_d}$ using the sequence 
\begin{align*}
	\Lambda^{(d-1)}_n(b_d)&:= \left(\Lambda^{(d)}_{b_d+\ell_d} \setminus\Lambda^{(d)}_{b_d}\right) \cap \{x_{d-1}<n-L\}\\
				& = \Lambda_L \cap \{ b_d\leq x_d<b_d + \ell_d \} \cap \{x_{d-1}<n-L\},
\end{align*}
which bounds the gap of each $\Lambda^{(d)}_{b_d+\ell_d} \setminus\Lambda^{(d)}_{b_d}$ by
\begin{align*}
	\gamma(\Lambda^{(d)}_{b_d+\ell_d} \setminus\Lambda^{(d)}_{b_d}) \geq \min_{\ell_{d-1} \leq n \leq 2L}\gamma(\Lambda^{(d-1)}_n(b_d)\setminus\Lambda^{(d-1)}_{n-\ell_{d-1}}(b_d)) \frac{(1-\epsilon_{d-1}\sqrt{\ell_{d-1}})^2}{\ell_{d-1}}.
\end{align*}
We iterate this process $d$ times: the sequence for $j$-th application martingale method  is
\begin{align*}
	\Lambda^{(j)}_n(b_{j+1}, \dots, b_d) &:= \left(\Lambda^{(j+1)}_{b_{j+1}+\ell_{j+1}} \setminus\Lambda^{(j+1)}_{b_{j+1}}\right) \cap \{x_{j}<n-L\}\\
				& = \Lambda_L \cap \{ \text{for } k>j, b_k\leq x_k<b_k + \ell_k \} \cap \{x_{j}<n-L\},
\end{align*}
for $\ell_j\leq n \leq 2L$.  
The last two sequences differ slightly.  The sequence in the $x_2$ direction is given by
\begin{align*}
	\Lambda^{(2)}_n(b_{3}, \dots, b_d) &:= \left(\Lambda^{(3)}_{b_{3}+\ell_{3}} \setminus\Lambda^{(3)}_{b_{3}}\right) \cap \{v\cdot x<n-L\}\\
				& = \Lambda_L \cap \{ \text{for } k>2, b_k\leq x_k<b_k + \ell_k \} \cap \{v\cdot x<n-L\},
\end{align*}
for $\ell_2 \leq n \leq 2L$.  For the $x_1$ direction, we use the sequence
\begin{align*}
	\Lambda^{(1)}_n(b_{2}, \dots, b_d) &:= \left(\Lambda^{(2)}_{b_{2}+\ell_{2}} \setminus\Lambda^{(2)}_{b_{2}}\right) \cap \{\frac{m}{m_1}\cdot x<n\}\\
				& = \Lambda_L \cap \{ \text{for } k>1, b_k\leq x_k<b_k + \ell_k \} \cap \{\frac{m}{m_1}\cdot x<n\},
\end{align*}
for $\ell_1 \leq n \leq L$.
  
Condition (ii) is satisfied by letting $\gamma_{\ell_j} := \min_n \gamma(\Lambda^{(j)}_n \setminus \Lambda^{(j)}_{n-\ell_j})$. 
To apply the Lemma \ref{lem:EpsilonCalculation} for Condition (iii), the volumes $\Lambda^{(j)}_n \setminus \Lambda^{(j)}_{n-\ell_j}$ intersected with the lattice must be connected.  
We verify this by checking that cross-sections of the volume are graph connected in each coordinate direction.  
The cross-sections of the volume parallel a general $(x_j,x_k)$ plane with $j,k \neq 1,2$ are rectangles and connected.  
The cross-sections with the $x_2$ coordinate and not the $x_1$ coordinate are either rectangles if $v_j =0$ or parallelograms with boundaries slope one which are connected since $L_2\geq \ell_2\geq 2$.  
For cross-sections with the $x_1$ coordinate, the argument in Case 1a applies under the condition $L_1\geq \ell_1 \geq \frac{m_j}{m_1} + 1$.  
We impose the condition $L \geq \ell_1 \geq \max_j(\frac{m_j}{m_1})+1$  to guarantees connectedness of the volumes.  

We now calculate $\epsilon_{\ell_j}$.  For the application in the $x_j$ direction, all terms independent of $x_j$ cancel in the ratio of normalization coefficients. We find
\begin{align*}
	\Vert G^{\Lambda^{(j)}_{n+1}\setminus\Lambda^{(j)}_{n+1-\ell_j}}E_n\Vert^2 \leq \max(\lambda_1^4, \, \lambda_1^{-4}) \cdot
	\frac{\tilde{\lambda}_j^{2(\ell_j-1)}(1-\tilde{\lambda}_j^{2(n+1-\ell_j)})(1-\tilde{\lambda}_j^2)}{
	(1-\tilde{\lambda}_j^{2\ell_2})(1-\tilde{\lambda}_j^{2n})},
\end{align*}
which is of the form stated in \eqref{CommonBound}. Therefore, we define
\begin{align*}
	\epsilon_{\ell_j} :=\max(\lambda_1^2, \, \lambda_1^{-2}) \min(1, \,
\tilde{\lambda}_j^{\ell_2-1}) \sqrt{\frac{1-\tilde{\lambda}_j^2
}{1-\tilde{\lambda}_j^{2\ell_j}}},
\end{align*}
for the smallest value of $\ell_j$ such that $\epsilon^2_{\ell_j} <\frac{1}{\ell_j}$.  
By the martingale method, we have
\begin{align*}
	\gamma(\Lambda^{(j)}_{2L}(b_{j+1}, \dots, b_d))\geq \min_n \gamma(\Lambda^{(j)}_n(b_{j+1}, \dots, b_d)\setminus\Lambda^{(j)}_{n-\ell_j}(b_{j+1}, \dots, b_d))\frac{(1-\epsilon_j\sqrt{\ell_j})^2}{\ell_j}.
\end{align*}

\textbf{Combining the Martingale Methods:}  
We have defined the sequences such that $\Lambda^{(j)}_{2L}(b_{j+1}, \dots, b_d)
= \Lambda^{(j+1)}_{b_{j+1}+\ell_{j+1}}(b_{j+2}, \dots, b_d)\setminus
\Lambda^{(j+1)}_{b_{j+1}}(b_{j+2}, \dots, b_d)$.
We iterate the method for each coordinate direction by applying the next
application to a general $ \gamma(\Lambda^{(j)}_n(b_{j+1}, \dots,
b_d)\setminus\Lambda^{(j)}_{n-\ell_j}(b_{j+1}, \dots, b_d))$.
Each application of the method results in a new constant factor and a minimum
over sets with fewer degrees of freedom. In the final application of the
martingale method, Condition (ii) requires a minimum over all $\gamma(\Lambda_n^{(1)}(b_2,\dots,b_d)\setminus
\Lambda_{n-\ell_1}^{(1)}(b_2,\dots,b_d))$, which are $d$-dimensional volumes
with coordinate lengths independent of the sequence. Let $\vec{\ell}$ denote the
vector $(\ell_1,\dots,\ell_d)$., and define
\begin{align*}
\Lambda_{\vec{\ell}} :=	\{x\in\mathbb{R}^d:\ 0\leq \frac{m}{m_1}\cdot x <\ell_1,\ 0\leq v\cdot x <\ell_2,\ \text{for } j\geq 3,\ 0\leq x_j<\ell_j\}.
\end{align*}
Then, the finite volumes for which we need to lower bound the gap, after a suitable translation in $\Ir^d$, are subsets of  $\Lambda_{\vec{\ell}} $.
We call the resulting set of translates $\mathcal{P(\vec{\ell})}$.
The family of sets $\mathcal{P(\vec{\ell})}$ itself is finite up to
translations. Therefore, combining bounds we obtain
\begin{align*}
	\gamma(\Lambda_L) \geq \min_{P\in\mathcal{P(\vec{\ell})}} \gamma(P) \prod_{j=1}^d \frac{(1-\epsilon_j\sqrt{\ell_j})^2}{\ell_j}>0.
\end{align*}


\noindent \textbf{Case 3b:} Suppose $\log\vec{\lambda} \neq
\pm\Vert\log\vec{\lambda}\Vert m$ and that there does not exist $k$ such that
$\lambda_k \neq 1$ and $m_k \neq 0$.

Recall that we are only considering $\vec{\lambda}\neq (1, 1, \ldots, 1)$.
Therefore, there is at least one $k$ such that $\lambda_k \neq 1$. Since $m$ is
a unit vector, not all $m_{1}, \dots, m_d$ are equal to zero. This implies there
must be at least one $\lambda_k =1$. We permute the variables so that $\lambda_1
= \ldots = \lambda_{j'-1}=1$, and $\lambda_{j'}, \lambda_{j'+1}, \dots,
\lambda_{d}$ are not equal to one. We assume, without loss of generality, that
$m_1= \max_j\{m_j\}$.

The proof for this case is essentially the same as the proof of Case 3a.  We
indicate how to choose the volumes $\Lambda_L$, how to obtain
$\epsilon_{\ell_j}$ and $\ell_1,\ldots, \ell_d$, and give adapted definitions
for $\tilde\lambda_j$, $\Lambda_{\vec\ell}$ and $\mathcal{P}(\vec\ell)$.

\textbf{Volumes:} The finite volumes for this case are defined as
\begin{align*}
	\Lambda_L = \{x\in\mathbb{Z}^d:&  0 \leq \frac {m}{m_1}\cdot x<  L ,\  -L\leq v\cdot x<  L,\ \text{ for } j=2,\dots, d-1,\ -L \leq x_j  <  L \},
\end{align*}
where $v = (-1,-2,\dots, -2, 0, \dots, 0, m_1)$, where $v_k=-2$ for
$k=2,\ldots, j'-1$. This vector determines two slanted boundaries which
guarantee the volumes satisfy the heuristic of Section \ref{sec:VolumeChoices}.

\textbf{Normalization Coefficient:}  
Define
\begin{displaymath}
   \tilde{\lambda}_j := \left\{
     \begin{array}{lr}
       \lambda_d & \text{ if } j=1 \\
       \lambda_d^{2-m_j/m_1}& \text{ if } 2\leq j \leq j'-1 \\
       \lambda_j & \text{ if } j' \leq j \leq d\\
     \end{array}
   \right.
\end{displaymath} 
Each modified parameter satisfies $\tilde\lambda_j\neq 1$. To see this,
note that by our original choice of ordering, $\lambda_j \neq 1$ for
$j = j',\ldots, d.$ Consequently, it also follows that
$\tilde\lambda_1=\lambda_d \neq 1$. Since $m_1 = \max_j\{m_j\}$, it follows that
$m_j/m_1\leq 1$, and therefore $\tilde\lambda_j = \lambda_d^{2-m_j/m_1}\neq 1$.

The general volume for all $d$ applications of the martingale method is of the
form
\begin{align*}
		\Lambda' = \{x\in\mathbb{Z}^d:& b_1 \leq  \frac {m}{m_1}\cdot x< b_1+  L_1,\  b_d \leq v\cdot x <  b_d + L_d \ \text{ for } j=2,\dots, d-1,\ b_j \leq x_j  < b_j + L_j \}.
\end{align*}
The compound inequality $b_1 \leq  \frac {m}{m_1}\cdot x< b_1+  L_1$ once again
bounds the values of $x_1$, where as the compound inequality $b_d \leq v\cdot x
< b_d + L_d$ bounds the values of $x_d$. To abbreviate notation, let $a_1(x) =
\sum_{j=2}^{j'-1}-\frac{m_jx_j}{m_1}+r$ and $a_d(x) =
\sum_{j=1}^{j'-1}(-v_j)x_j$. The remainder term for $a_1(x)$ only depends
on $x_2, \dots, x_{j'-1}$. The normalization coefficient is then
\begin{align*}
C(\Lambda') &= \sum_{x_{d-1}=b_{d-1}}^{b_{d-1}+L_{d-1} - 1}  \lambda_{d-1}^{2x_{d-1}} \dots\sum_{x_2=b_2}^{b_2+L_2- 1}  \lambda_2^{2x_2}\sum_{x_1=b_1 + a_1(x)}^{b_1+L_1 -1+ a_1(x)} \lambda_1^{2x_1}\sum_{x_d=b_d + a_d(x) }^{b_d+L_d - 1+ a_d(x)}  \lambda_d^{2x_d}\\
			&= \left(\prod_{j=1}^d \tilde\lambda_j^{2b_j}  \right)\sum_{x_d=0}^{L_d - 1} 
			\tilde{\lambda}_d^{2x_d}\sum_{x_{d-1}=0}^{L_{d-1} - 1}  \tilde{\lambda}_{d-1}^{2x_{d-1}} \dots\sum_{x_2=0}^{L_2- 1}  \tilde{\lambda}_2^{2x_2}\lambda_1^{2r}\sum_{x_1=0}^{L_1 -1} \tilde{\lambda}_1^{2x_1}.
\end{align*}

\textbf{Martingale Method:}  
For brevity, we define the $\epsilon_{\ell_j}$ and $\ell_j$ and refer to Case 3a for the bound. 
We define
\begin{align*}
	\epsilon_{\ell_j} :=\max(\lambda_d^2, \lambda_d^{-2}) \min(1, \,
\tilde{\lambda}_j^{\ell_2-1}) \sqrt{\frac{1-\tilde{\lambda}_j^2
}{1-\tilde{\lambda}_j^{2\ell_j}}}.
\end{align*}
for the smallest value of $\ell_j$ such that $\epsilon^2_{\ell_j}
<\frac{1}{\ell_j}$.
We may drop the error term $\max(\lambda_d^2, \lambda_d^{-2})$ for all $j\geq
j'$ since $m_j = 0$ implies $r$ does not depend $x_j$. Analogous to previous
case, we define $\Lambda_{\vec\ell}$ as
\begin{align*}
\Lambda_{\vec{\ell}} :=\{x\in\mathbb{R}^d:\ 0\leq \frac{m}{m_1}\cdot x <\ell_1,\ 0\leq v\cdot x <\ell_d,\ \text{for } 2\leq j\geq d-1,\ 0\leq x_j<\ell_j\}.
\end{align*}
Once again, the finite volumes for which we need to lower bound the spectral
gap, after a suitable translation in $\Ir^d$, are subsets of 
$\Lambda_{\vec{\ell}}$. Let $\mathcal{P(\vec{\ell})}$ denote the resulting
finite set of translates. After $d$ iterations of the martingale method, as in
Case 3a, we have a positive lower bound on the gap of the form
\begin{align*}
	\gamma(\Lambda_L) \geq \min_{P\in\mathcal{P}(\vec{\ell})}\gamma(P) \prod_{j=1}^d\frac{(1-\epsilon_j\sqrt{\ell_j})^2}{\ell_j}.
\end{align*}
We note that the parameters $\ell_j,\epsilon_j$, and the set
$\mathcal{P}(\vec{\ell})$ are different from those in Case 3a. Once again using
Theorem \ref{thm:GNSSpecGap} we conclude that $\gamma(D)>0$.

\noindent \textbf{Case 4}:  Suppose $\log\vec{\lambda} = \Vert\log\vec{\lambda}\Vert m$.

As described in Remark \ref{rem:coordinates} we apply coordinate reflections so
each $m_j \geq 0$ and permute the indices so that $m_1>0$. We proceed by
stating how to choose the volumes $\Lambda_L$, how to obtain good values for
 $\ell_1,\ldots, \ell_d$, and give adapted definitions for $\tilde\lambda_j$,
 $j=1,\ldots,, d$, and two analogues of $\Lambda_{\vec\ell}$ and
 $\mathcal{P}(\vec\ell)$.

 \textbf{Volumes:} 
 We define a vector $v := (1, v_2, \dots, v_d)$ with 
 \begin{displaymath}
   v_j= \left\{
     \begin{array}{lr}
       -1 & : m_j = 0\\
       0 & : m_j \neq 0
     \end{array}
	\right.
\end{displaymath} 
For the sequence $\Lambda_L$ we take
\begin{align*}
	\Lambda_L:= \{ x \in \mathbb{Z}^d: \, & \frac{m\cdot x}{m_1}>0,\ v\cdot x< L+
	\sum_{j=2}^d (\frac{m_j}{m_1} - v_j)L, \text{ for } j=2,\dots, d,\ -L \leq x_j  < L\}.
\end{align*}
The summation term on the bound of $v\cdot x$ is used to guarantee a length of
at least $L$ in the $x_1$ direction for any $x_2,\dots,x_d$ in $[-L,L)$.

\textbf{Normalization Coefficients for $x_2, \dots, x_d$ directions:}
To simplify notation, we define $\tilde{\lambda}_j$ as
\begin{align*}
	\tilde{\lambda}_1 := \lambda_1,\quad \text{and} \quad
	\tilde{\lambda}_j := \lambda_1^{\frac{m_j}{m_1} -v_j} \quad \text{for}
	\quad j\geq 2,
\end{align*}
which are all strictly greater than one since $\log\lambda_j =
\Vert\log\vec{\lambda}\Vert m_j$ and $m_j\geq 0$. The general volume $\Lambda'$
defined below covers the volumes generated by the martingale method for the $x_2, \dots, x_d$ directions.
\begin{align*}
	\Lambda' := \{ x\in\mathbb{Z}^d: \,& 0\leq \frac{m\cdot x}{m_1},\ v\cdot x <
	L_1 + \sum_{j=2}^d (\frac{m_j}{m_1} - v_j)L,\  for\ j=2,\dots,d,\ b_j \leq x_j < b_j+L_j\},
\end{align*}
where $b_j \geq -L$ and $L_j \leq 2L$.  
Define 
$$a_1(x) = \sum_{j=2}^d\tfrac{-m_jx_j}{m_1}, \quad \text{and} \quad b_1(x) =
\sum_{j=2}^d (\tfrac{m_j}{m_1}L-v_j(x_j+L)).$$ 
Furthermore, let $r(x)$ be the remainder associated with $a_1(x)$, that is $r(x)
= \lceil a_1(x) \rceil -a_1(x)$, and define $r(L)$ be the remainder term
associated with $x_2=x_3=\dots = x_d = -L$. The normalization constant is given by
\begin{align*}
	C(\Lambda') 
		&= \sum_{x_d = b_d}^{b_d+ L_d -1} \lambda_d^{2x_d} \dots
		\sum_{x_2=b_2}^{b_2+L_2-1} \lambda_2^{2x_2}\sum_{x_1= a_1(x) + r(x)}^{L_1 -1
		+ b_1(x) + r(L)}\lambda_1^{2x_1}\\
		&= \sum_{x_d = b_d}^{b_d+ L_d -1}
		(\lambda_1^{\frac{-m_d}{m_1}}\lambda_d)^{2x_d} \dots
		\sum_{x_2=b_2}^{b_2+L_2-1} 
		(\lambda_1^{\frac{-m_2}{m_1}}\lambda_2)^{2x_2}\lambda_1^{2r(x)}\sum_{x_1=
		0}^{L_1 -1 +b_1(x) -a_1(x) + r(L)-r(x)}\lambda_1^{2x_1}\\
		&= \sum_{x_d = b_d}^{b_d+ L_d -1} \dots
		\sum_{x_2=b_2}^{b_2+L_2-1}\lambda_1^{2r(x)} \frac{ \lambda_1^{2(L_1
		+b_1(x)-a_1(x)+ r(L)-r(x))}-1}{\lambda_1^2 -1}.
\end{align*}
In the above computation, we use that $\lambda_1^{\frac{-m_j}{m_1}}\lambda_j =1$
as $\log\vec{\lambda}=\Vert\log\vec{\lambda}\Vert m$. Just as in the proof of
Case 2a, the exponent of $\lambda_1$ is greater than $2$ since
$b_1(x)-a_1(x)\geq 0$, $r(x)<1$, and we require that $L_1\geq\ell_1\geq 2$.
Therefore, the bounds from Case 2a hold here, that is
\begin{align*}
\frac{ \lambda_1^{2(L_1 +b_1(x)-a_1(x)+ r(L)-r(x))}-1}{\lambda_1^2 -1} &\geq
C_{min} \lambda_1^{2(L_1 +b_1(x)-a_1(x)+ r(L)-r(x))}\\
			\frac{ \lambda_1^{2(L_1 +b_1(x)-a_1(x)+ r(L)-r(x))}-1}{\lambda_1^2 -1} &\leq
			C_{max} \lambda_1^{2(L_1 +b_1(x)-a_1(x)+ r(L)-r(x))}
\end{align*}
Therefore, $C(\Lambda')$ is bounded above and below by
\begin{align*}
	C_{min}\tilde{\lambda}_1^{2(L_1 +
    r(L))}\prod_{j=2}^d\left(\tilde\lambda_j^{2(b_j+L)}\sum_{x_j = 0}^{L_j -1}
    \tilde{\lambda}_j^{2(x_j)} \right)
	\leq C(\Lambda') \leq 
    C_{max}\tilde{\lambda}_1^{2(L_1 +
    r(L))}\prod_{j=2}^d\left(\tilde\lambda_j^{2(b_j+L)}\sum_{x_j = 0}^{L_j -1}
    \tilde{\lambda}_j^{2(x_j)} \right).
\end{align*}

\textbf{Martingale Method in $x_2,\dots,x_d$ Directions:}
The first $d-1$ applications of the martingale method follow the same
procedure, which we now detail. The last application will be slightly
different and will need to be considered separately. We apply the martingale
method first to the sequence
\begin{align*}
	\Lambda^{(d)}_n := \Lambda_L \cap \{ x_d<n-L\}, \quad 0\leq n \leq 2L.
\end{align*}
Replacing $n-L$ by $b_d+\ell_d$, we apply the martingale method to each $\Lambda^{(d)}_{b_d+\ell_d}\setminus\Lambda^{(d)}_{b_d}$ using the sequence
\begin{align*}
	\Lambda^{(d-1)}_n(b_d) =\left( \Lambda^{(d)}_{b_d+\ell_d}\setminus\Lambda^{(d)}_{b_d}\right) \cap \{x_{d-1} < -L + n \}, \quad 0\leq n \leq 2L.  
\end{align*}
We iterate this procedure for $j\geq 2$, applying the martingale method to the sequence $\Lambda^{({j+1})}_{b_{j+1}+\ell_{j+1}}\setminus\Lambda^{({j+1})}_{b_{j+1}}$ using the sequence
\begin{align*}
	\Lambda^{(j)}_n(b_{j+1}, \dots, b_d) =\left( \Lambda^{({j+1})}_{b_{j+1}+\ell_{j+1}}(b_{j+2},\dots, b_d)\setminus\Lambda^{({j+1})}_{b_{j+1}}(b_{j+2},\dots, b_d)\right) \cap \{x_{j} < -L + n \}
\end{align*}

Condition (ii) is satisfied by taking the minimum over 
$\Lambda^{(j)}_n \setminus \Lambda^{(j)}_{n-\ell_j}$.
 The volumes are connected and we may apply Lemma \ref{lem:EpsilonCalculation} and compute the norm
 \begin{equation}
\|G^{\Lambda^{(j)}_{n+1}\setminus\Lambda^{(j)}_{n+1-\ell}}E_n\|^2 \leq 
	\frac{C_{max}^2}{C_{min}^2}\cdot
	\frac{1-\tilde{\lambda}_j^{2(n+1-\ell_j)}}{1-\tilde{\lambda}_j^{2n}} \cdot
	\frac{\tilde{\lambda}_j^{2(\ell_j-1)}(1-\tilde{\lambda}_j^2)}{1-\tilde{\lambda}_j^{2\ell_2}}.
\end{equation}
This is of the form given in \eqref{CommonBound}, so we choose
\begin{equation}
	\epsilon_{\ell_j} := \frac{C_{max}}{C_{min}}\sqrt{\frac{1-\tilde{\lambda}_j^2}{1-\tilde{\lambda}_j^{2\ell_j}}},
\end{equation} 
for the smallest value of $\ell_j\geq 2$ such that
$\epsilon_{\ell_j}^2<\frac{1}{\ell_j}$.
For each $j\neq 1$ the martingale method in the $x_j$ direction gives a bound of the form
\begin{equation}
	\gamma(\Lambda^{(j)}_{2L}(b_{j+1}, \dots, b_d))\geq \min_{b_j} \gamma(\Lambda^{({j})}_{b_{j}+\ell_{j}}(b_{j+1},\dots, b_d)\setminus\Lambda^{({j})}_{b_{j}}(b_{j+1},\dots, b_d))\frac{(1-\epsilon_{\ell_j}\sqrt{\ell_j})^2}{\ell_j}.
\end{equation}
 
 \textbf{Martingale Method in the $x_1$ Direction:}
 For the last application of the martingale method to $\Lambda^{(2)}_{b_2+\ell_2}(b_{3}, \dots, b_d)\setminus \Lambda^{(2)}_{b_2}(b_{3}, \dots, b_d)$, we use the sequence
\begin{align*}
	\Lambda^{(1)}_n(b_{2}, \dots, b_d)
	&=\left(\Lambda^{({2})}_{b_{2}+\ell_{2}}(b_{3},\dots,b_d)\setminus\Lambda^{({2})}_{b_{2}}(b_{3},\dots, b_d)\right) \cap \{v\cdot x <  n + \sum_{j=2}^d (\frac{m_j}{m_1} - v_j)(-b_j)\}\\
	&=\{ x\in\bZ^d : \, 0 \leq \frac{m\cdot x}{m_1}, \, v\cdot x < n +
	\sum_{j=2}^d (\frac{m_j}{m_1} - v_j)(-b_j),\  for\ j=2,\dots,d,\ b_j \leq x_j <
	b_j+\ell_j\}.
\end{align*}
The form of the equations from this set are the same as those from $\Lambda'$
where we have replaced $L_1$ with $n$, and $L$ with $-b_j$. As such, the upper
and lower bounds from $C(\Lambda')$ hold for $C(\Lambda_n^{(1)}(b_2,\ldots,
b_d))$ after making the appropriate substitutions. We will use this fact when
discussing Condition (iii). We note that this volume has length $n$ in the the
$x_1$ direction when $x_j=b_j$ for all $j\neq 1$, and as such the bounds on
$x_1$ are well defined.

For Condition (ii) we need to consider the minimum of the spectral
gaps of two finite families of volumes. 
For $n>\ell_1$, the volumes have the form 
 \begin{align*}
	&\Lambda^{(1)}_n(b_{2}, \dots, b_d)\setminus \Lambda^{(1)}_{n-\ell_1}(b_{2}, \dots, b_d)\\ 
	&:= \{ x\in\mathbb{Z}^d: \text{ for } j=2,\dots,d,\ b_j \leq x_j < b_j+\ell_j,
	n-\ell_1\leq v\cdot x + \sum_{j=2}^d (\frac{m_j}{m_1} - v_j)b_j< n\}.
\end{align*}
These volumes are isomorphic for any choice of $n, b_2, \dots, b_d$ since the
slanted boundary generated by $v$ only has nonzero components $1$ and $-1$. We
denote by $P(\vec{\ell})$ a representative of these volumes.
For $n=\ell_1$, $\Lambda^{(1)}_{n-\ell_1} = \Lambda^{(1)}_{0} = \emptyset$ by
convention. In this case we are left to consider
 \begin{align*}
	&\Lambda^{(1)}_{\ell_1}(b_{2}, \dots, b_d)\\
	& := \{ x\in\mathbb{Z}^d:0 \leq \frac{m\cdot x}{m_1},\ v\cdot x < \ell_1 + \sum_{j=2}^d (\frac{m_j}{m_1} - v_j)(-b_j),
	  \text{ for }\ j=2,\dots,d,\ b_j \leq x_j < b_j+\ell_j\}.
\end{align*}
 We denote the resulting finite family of volumes as $\mathcal{T}(\vec{\ell})$.
 Condition (ii) is satisfied by taking a minimum of spectral gaps over the union of both sets
 \begin{equation}
 	\gamma_{\ell_1} : = \min_{\Lambda \in \{P(\vec{\ell})\}\cup\mathcal{T}(\vec{\ell})} \gamma(\Lambda).
 \end{equation}
 
For Condition (iii), the volumes are connected and we may apply Lemma
\ref{lem:EpsilonCalculation}. To simplify the bound, recall that
$$\frac{C(\Lambda_{n+1}^{(1)}\backslash\Lambda_n^{(1)})}{C(\Lambda_{n+1}^{(1)}
\backslash\Lambda_{n+1-\ell_2}^{(1)})} \leq 1.$$ 
 Using the bounds on $C(\Lambda_n^{(1)})$ as previously stated, we bound the norm
 by
 \begin{equation}
\|G^{\Lambda^{(1)}_{n+1}\setminus \Lambda^{(1)}_{n+1-\ell}}E_n\|^2 \leq \frac{C(\Lambda^{(1)}_{n+1-\ell_1})}{C(\Lambda^{(1)}_{n})} \leq 
\frac{C_{max}}{C_{min}}\cdot \tilde{\lambda}_1^{-2(\ell_1-1)},
\end{equation}
which decays exponentially as $\tilde{\lambda}_1>1$. We choose
\begin{equation}
	\epsilon_{\ell_1} := \max(\lambda_1, \lambda_1^{-1}) \tilde{\lambda}_1^{-(\ell_1-1)},
\end{equation} 
 for the smallest value of $\ell_1\geq 2$ such that 
 $\epsilon_{\ell_1}^2<\frac{1}{\ell_1}$.
 
 \textbf{Combining the Bounds:}
 As in Case 3a, we iterate the bounds in each direction.  
 The bound on $\gamma(\Lambda_L)$ is
 \begin{equation}
 	\gamma(\Lambda_L) \geq  \min_{\Lambda \in \{P(\vec{\ell})\}\cup\mathcal{T}(\vec{\ell})} \gamma(\Lambda) \prod_{j=1}^d \frac{(1-\epsilon_{\ell_j}\sqrt{\ell_j})^2}{\ell_j}>0.
 \end{equation}
 Using  Theorem \ref{thm:GNSSpecGap}, this is a positive lower bound for $\gamma(D)$.  
 This completes the proof of the lower bounds for all cases.
 
\textbf{Sketch of Proof for Corollary \ref{cor:GappedZd}:}  Suppose
$\log\vec{\lambda}\neq 0$.  Choose a unit vector $m$ such that
$\log\vec{\lambda} \neq \pm \Vert\log\vec{\lambda}\Vert m$ and for which there
exists $j\in\{1,\dots,d\}$ such that $\lambda_j \neq 1$ and $m_j \neq 0$ Then,
the conditions of Case 3a are satisfied. Since the PVBS models are translation
invariant, the sequence of volumes $\Lambda_L$ from Case 3a may be translated in
such a way that $\Lambda_L\nearrow\bZ^d$. Therefore, by Theorem
\ref{thm:GNSSpecGap} $$\gamma(\bZ^d)\geq \liminf \gamma(\Lambda_L)>0.$$

\section{The Upper Bound on the Spectral Gap}\label{sec:upperbound}

Recall that $\caG^D$ is one-dimensional and spanned by $\Omega_0^D$. It follows
that any one particle state in the GNS Hilbert space $\caH_D$ is orthogonal to
the ground state space.
By the variational principle the energy of any state in $\cH_D$ that is
orthogonal to $\caG_D$ is therefore an upper bound for the spectral gap. That
is, if $\Psi\perp \caG_D$, then 
\[\gamma(D) \leq \frac{\braket{\Psi}{H^D\Psi}}{\|\Psi\|^2}.\]
We prove Theorem \ref{thm:upperbound} by applying the variational principle to
the sequence of one particle states, $\Psi^L$, of the form
\begin{align*}
\Psi^L = \sum_{x\in \Lambda^L} \lambda^{x} \pi_0(\sigma_{x}^1)\Omega_0^D
\end{align*}
where $\pi_0$ is the GNS representation and $\Lambda_L$ is defined by
\begin{align*}
	\Lambda_L :=\{x\in\mathbb{Z}^d: 0\leq m\cdot x < L,\ \text{ for } j=2,\dots, d,\ -L \leq x_j \leq L\} .
\end{align*}
Note that $\pi_0(\sigma_{x}^1)\Omega_0$ is the product state describing a
particle localized at the site $x$, and that $\Lambda_L\nearrow D$. 
For any observable $A$ with finite support, the GNS representative $\pi_0(A)$ 
acts on the product vector $\Omega_0^D$ in the canonical way. From now on, 
we will drop $\pi_0$ in the notation.

Theorem \ref{thm:upperbound} is proved by establishing that for all $d\geq
2$, $\vec\lambda\in (0,\infty)^d$, and unit vectors $m\in\Rl^d$,
\be\label{upperclaim}
\limsup_{L\to\infty}\frac{\langle\Psi^L, H^D\Psi^L\rangle}{\|\Psi^L\|^2}
\leq \frac{2(d-1)}{c(m) c(\vec\lambda)^2}\|\log\vec{\lambda}\||\sin(\theta)|,
\ee
where $\theta$ is the angle between $\log\vec{\lambda}$ and the outward normal
$-m$ of the hyperplane $D$, and $c(\cdot)$ is defined as in \eq{cv}.
We restrict to the case that $\theta \in (-\pi/2,\pi/2)$ as for all other values
of $\theta$, the lower bounds derived in Sections \ref{sec:TwoDimCase} and
\ref{sec:dDimCase} show that the gap does not vanish, and we are interested in
the behavior near $\theta = 0$.

\begin{proof}
The expression for the upper bound given in the theorem is invariant under the
permutations and reflections.
We use the coordinate transformations discussed in Remark \ref{rem:coordinates}
to assume $m_j \geq 0$ for all $j$. Since we assume $m\cdot \log \vec\lambda<0$,
there must be at least one value of the index $j$ such that $m_j>0$ and
$\lambda_j\in (0,1)$. Without loss of generality we assume this holds for $j=1$.

We prove a lower bound on $\|\Psi^L\|^2$ and an upperbound on
$\braket{\Psi^L}{H^D\Psi^L}$ to obtain \eqref{upperclaim}. Let
$\tilde\lambda_j~=~\lambda_1^{-m_j/m_1}\lambda_j$ for $j \neq 1$.
For a fixed $x_2, \dots, x_d$ in $\Lambda_L$, we denote the minimum value of $x_1$ by
$a(x) = -\sum_{j=2}^d\frac{-m_jx_j}{m_1} + r$, where $r<1$ is a remainder term
that depends on $x_2, \ldots, x_d$. Since $\lambda_1<1$, it follows that
$\lambda_1^{2r}\geq \lambda_1^2$, and we find
\begin{align}
	 \Vert \Psi^L\Vert^2&= \sum_{x_d=-L}^{L}\lambda_d^{2x_d} \dots \sum_{x_2=-L}^{L} \lambda_2^{2x_2}
	 \sum_{x_1=a(x)}^{a(x)+ L-1} \lambda_1^{2x_1}\nonumber\\
	 	&\geq \lambda_1^{2}\sum_{x_d=-L}^{L}\tilde\lambda_d^{2x_d}\dots
	 	\sum_{x_2=-L}^{L} \tilde\lambda_2^{2x_2}\sum_{x_1=0}^{ L-1} \lambda_1^{2x_1}
	 	\nonumber \\
		&=
		\lambda_1^2\left(\frac{1-\lambda_1^{2L}}{1-\lambda_1^2}\right)\prod_{j=2}^d\left(\sum_{x_j=-L}^L
		\tilde\lambda_j^{2x_j}\right).\label{TotalLowerBound}
\end{align}
We note that $C(\Lambda_L) = \|\Psi^L\|^2$ diverges as $\Lambda_L\nearrow D$
since each sum $\sum_{x_j=-L}^{L}\tilde\lambda_j^{2x_j}$ diverges. Appealing to
\newline \cite[Proposition 2.2]{bachmann:2015}, shows the lower bound on
$\|\Psi^L\|^2$ explicitly proves that $\caG_D = \text{span}\{\Omega_0^D\}$.

For the upperbound on $\braket{\Psi^L}{H^D\Psi^L}$, since
\[H^D = \sum_{j=1}^d\sum_{x, x+e_j\in
D}h_{x,x+e_j}\]
and $\Psi^L$ is the one particle ground state on $\Lambda_L$ and the
zero particle ground state on $D\setminus\Lambda_L$, the only nonzero
contributions to the energy are given by interactions with support $\{x, \, y\}$
such that $x\in \Lambda_L$ and $y\in D\setminus\Lambda_L$, see Figure
\ref{figure:upperbound}.
Since the PVBS model is nearest neighbor, it follows that there exists a $j$
such that either $y=x+e_j$ or $y = x - e_j$. Therefore,
\begin{equation*}
\braket{\Psi^L}{H^D\Psi^L} = \langle\Psi^L, \,
\hskip -5pt \sum_{\substack{x\in\Lambda_L, y\notin\Lambda_L \\
|x-y|=1}}\hskip -5pt h_{x,y}\Psi^L\rangle
\end{equation*}
\begin{figure}
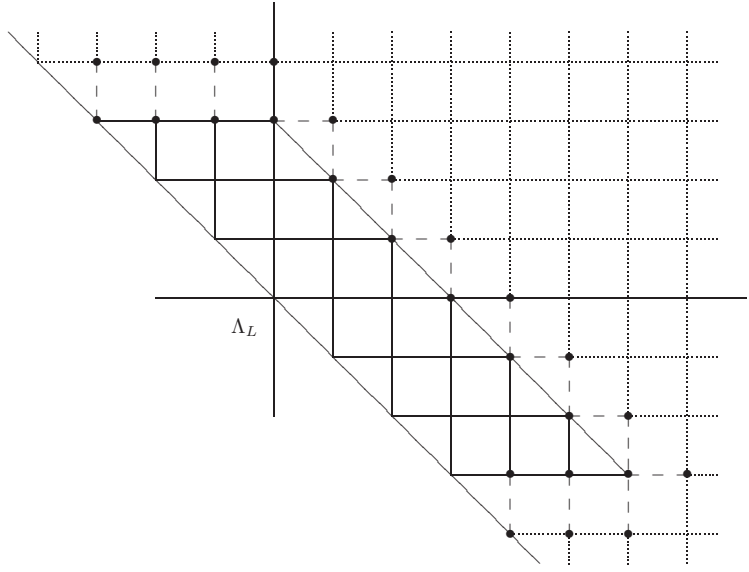


 \begin{adjustbox}{max size={.6\textwidth}{.5\textheight}}
{
\xy
(45,-45); (-45,45) **\dir{-};
(-5,-5)*+{\Lambda_L};
(-30,40)*+{\bullet};
(-20,40)*+{\bullet};
(-10,40)*+{\bullet};
(0,40)*+{\bullet};
(-30,30)*+{\bullet};
(-20,30)*+{\bullet};
(-10,30)*+{\bullet};
(0,30)*+{\bullet};
(60,-40)*+{\bullet};
(50,-40)*+{\bullet};
(40,-40)*+{\bullet};
(60,-30)*+{\bullet};
(50,-30)*+{\bullet};
(40,-30)*+{\bullet};
(70,-30)*+{\bullet};
(50,-20)*+{\bullet};
(60,-20)*+{\bullet};
(40,-10)*+{\bullet};
(50,-10)*+{\bullet};
(30,0)*+{\bullet};
(40,0)*+{\bullet};
(20,10)*+{\bullet};
(30,10)*+{\bullet};
(10,20)*+{\bullet};
(20,20)*+{\bullet};
(10,30)*+{\bullet};
(-30,40);(-30,30)**\dir{--};
(-20,40);(-20,30)**\dir{--};
(-10,40);(-10,30)**\dir{--};
(0,40);(0,30)**\dir{--};
(0,30);(10,30)**\dir{--};
(10,20);(10,30)**\dir{--};
(10,20);(20,20)**\dir{--};
(20,10);(20,20)**\dir{--};
(30,10);(30,0)**\dir{--};
(20,10);(30,10)**\dir{--};
(30,0);(40,0)**\dir{--};
(40,0);(40,-10)**\dir{--};
(40,-10);(50,-10)**\dir{--};
(50,-10);(50,-20)**\dir{--};
(50,-20);(60,-20)**\dir{--};
(60,-20);(60,-40)**\dir{--};
(60,-30);(70,-30)**\dir{--};
(50,-30);(50,-40)**\dir{--};
(40,-30);(40,-40)**\dir{--};
(-30,30); (0,30)**\dir{-}; 
(0,30); (60,-30)**\dir{-}; 
(30,-30); (60,-30)**\dir{-};
(-20,20); (10,20)**\dir{-};
(-10,10); (20,10)**\dir{-};
(10,-10); (40,-10)**\dir{-};
(20,-20); (50,-20)**\dir{-};
(-20,30);(-20,20)**\dir{-};
(-10,30);(-10,10)**\dir{-};
(10,20);(10,-10)**\dir{-};
(20,10);(20,-20)**\dir{-};
(30,0);(30,-30)**\dir{-};
(40,-10);(40,-30)**\dir{-};
(50,-20);(50,-30)**\dir{-};
(0,-20);  (0,50) **\dir{-};
(-10,40); (-10,45) **\dir{.};
(-20,40);  (-20,45) **\dir{.};
(-30,40);  (-30,45) **\dir{.};
(-40,40);  (-40,45) **\dir{.};
(10,30);  (10,45) **\dir{.};
(20,20);  (20,45) **\dir{.};
(30,10);  (30,45) **\dir{.};
(40, 0);  (40,45) **\dir{.};
(50,-45);  (50,-40) **\dir{.};
(50,-10);  (50,45) **\dir{.};
(60,-45);  (60,-40) **\dir{.};
(60,-20);  (60,45) **\dir{.};
(70,-45);  (70,45) **\dir{.};
(-20,0);  (80,0) **\dir{-};
(30,10);  (75,10) **\dir{.};
(20,20);  (75,20) **\dir{.};
(10,30);  (75,30) **\dir{.};
(-40,40); (75,40) **\dir{.};
(0,0);  (75,0) **\dir{.};
(50,-10); (75,-10) **\dir{.};
(60,-20); (75,-20) **\dir{.};
(70,-30);(75,-30)**\dir{.};
(40,-40); (75,-40) ** \dir{.};
\endxy 
}
\end{adjustbox}
   \caption{A typical $\Lambda_L$ in $d=2$. An edge between a site on
   the boundary of $\Lambda_L$ and a site outside of $\Lambda_L$
   indicates a nearest neighbor interaction that contributes nonzero energy to
   $\braket{\Psi^L}{H^D\Psi^L}$.}
   \label{figure:upperbound}
\end{figure}
Note that for a site $x\in\Lambda_L$ to have a nearest neighbor outside
$\Lambda_L$ requires that $x$ is on the boundary of $\Lambda_L$. Computing
$\braket{\Psi^L}{h_{x,y}\Psi^L}$ for $x$ on the boundary and $y
= x+e_j$ or $y = x-e_j$ outside of $\Lambda_L$ produces
\begin{align*}
\braket{\Psi^L}{h_{x,x+e_j}\Psi^L} & =
\lambda^{2x}\frac{\lambda_j^2}{\lambda_j^2+1}\leq \lambda^{2x} \\
\braket{\Psi^L}{h_{x-e_j,x}\Psi^L} & =
\lambda^{2x}\frac{1}{\lambda_j^2+1}\leq \lambda^{2x}
\end{align*}
We consider each boundary that is connected to $D\setminus\Lambda_L$ separately.
For $j\geq 2$ and $x\in\Lambda_L$ such that $x_j = L$, the only nearest neighbor
$y\in D\setminus\Lambda_L$ is $y = x+e_j$. Similarly, for $x\in \Lambda_L$ such
that $x_j = -L$, the only nearest neighbor $y\in D\setminus\Lambda_L$ is $y = x-e_j$.
For $x\in\Lambda_L$ such that $x_1 = a(x)+L-1$, there is a maximum of $d$
nearest neighbors $y\in D\setminus\Lambda_L$, namely $y_1 = x+e_1, \ldots,
y_d=x+e_d,$ see Figure \ref{figure:upperbound}. Since the interaction terms of
the PVBS Hamiltonian are non-negative, it follows that
\begin{align}
\braket{\Psi^L}{H^D\Psi^L} \quad \leq & \quad \sum_{j=2}^d
\sum_{\substack{x\in\Lambda_L\\x_j = L}}
		\braket{\Psi^L}{h_{x,x+e_j}\Psi^L}
		 \quad + \quad \sum_{j=2}^d \sum_{\substack{x\in\Lambda_L\\x_j = - L}}
		\braket{\Psi^L}{h_{x-e_j, x}\Psi^L} \nonumber\\
		& \hskip35pt + \hskip-10pt\sum_{\substack{x\in\Lambda_L \\ x_1 =
		a(x)+L-1}}\sum_{j=1}^d \braket{\Psi^L}{h_{x,x+e_j}\Psi^L} \nonumber\\
		\leq & \quad \sum_{j=2}^d\sum_{\substack{x\in\Lambda_L \\ x_j = \pm
		L}}\lambda^{2x} \quad + \quad d\hskip-15pt\sum_{\substack{x\in \Lambda_L
		\\ x_1 = a(x)+L-1}}\lambda^{2x} \label{TotalUpperBound}
\end{align}
For a fixed value of $j = 2, \ldots d$, we have
\begin{align}
      \sum_{\substack{x\in\Lambda_L \\ x_j = \pm L}}\lambda^{2x}& = \sum_{x_j =
      \pm L}\ \sum_{x_d=-L}^{L}\lambda_d^{2x_d} \dots \sum_{x_2=-L}^{L}
      \lambda_2^{2x_2}\sum_{x_1=a(x)}^{a(x)+ L-1} \lambda_1^{2x_1}\nonumber \\
	&\leq \left(\tilde\lambda_j^{2L}+\tilde\lambda_j^{-2L}\right)\left(\prod_{k\neq
	1,j}\sum_{x_k=-L}^{L}
	\tilde\lambda_k^{2x_k}\right)\left(\frac{1-\lambda_1^{2L}}{1-\lambda_1^2}\right)\label{FirstPartUpperBound}
\end{align} 
where we use $\lambda_1^{2r}\leq 1$. Similarly,
 \begin{align}
 	 \sum_{\substack{x\in \Lambda_L \\ x_1 = a(x)+L-1}}\lambda^{2x}& =
 	 \sum_{x_d=-L}^{L}\lambda_d^{2x_d} \dots \sum_{x_2=-L}^{L} \lambda_2^{2x_2} \lambda_1^{2(a(x) + L-1)}\nonumber\\
 	 &\leq \lambda_1^{2(L-1)}\left(\prod_{k=2}^d\sum_{x_k=-L}^{L}
 	 \tilde\lambda_k^{2x_k}\right)\label{SecondPartUpperBound}
\end{align}
We replace \eqref{FirstPartUpperBound} and \eqref{SecondPartUpperBound} into
\eqref{TotalUpperBound} to obtain the final upper bound for
$\braket{\Psi^L}{H^D\Psi^L}$. Using the lower bound from
\eqref{TotalLowerBound}, we find that the overall energy contribution is bounded
above by
\begin{align}
	\frac{\langle\Psi^L, H^D\Psi^L\rangle}{\langle \Psi^L, \Psi^L\rangle}
	&\leq
	\lambda_1^{-2}
	\left[\frac{d\lambda_1^{2(L-1)}(1-\lambda_1^2)}{1-\lambda_1^{2L}} +
	\sum_{j=2}^d\frac{\left(\tilde\lambda_j^{2L}
	+ \tilde\lambda_j^{-2L}\right)}{\left(\sum_{x_j=-L}^L
	\tilde\lambda_j^{2x_j}\right)}\right] \label{GapUpperBound}
\end{align} 

The first term is exponentially small in $L$, and tends to zero. For each
$j>1$, if $\tilde\lambda_j = 1$, then the corresponding term in the sum equals
$2/(2L+1)$ and converges to zero as $L$ goes to infinty. If $\tilde{\lambda}_j
\neq 1$, treating the cases $\tilde\lambda_j<1$ and $\tilde\lambda_j>1$
separately at taking the limit $L\to\infty$ we find
\begin{align*}
	\frac{\tilde\lambda_j^{2L}+\tilde\lambda_j^{-2L}}{\sum_{x_j=-L}^L
	\tilde\lambda_j^{2x_j}} \rightarrow 1 - \min(\tilde\lambda_j, \,
	\tilde\lambda_j^{-1})^2.
\end{align*}
Using the definition $\tilde\lambda_j = \lambda_j\lambda_1^{-m_j/m_1},$ we can
rewrite $\min(\tilde\lambda_j, \, \tilde\lambda_j^{-1})^2 =
e^{-2|\log\lambda_j - \tfrac{m_j}{m_1}\log\lambda_1|}$. Note that in the case
that $\tilde\lambda_j = 1$, we have $0 = 1- \min(\tilde\lambda_j, \,
\tilde\lambda_j^{-1})^2$. Therefore, by the variational principle 
\begin{align*}\gamma(D) & \leq \lambda_1^{-2}\sum_{j=2}^d 1- e^{-2|\log\lambda_j -
\tfrac{m_j}{m_1}\log\lambda_1|} \\
	& \leq \frac{2}{m_1 \lambda_1^2}\sum_{j=2}^d |- m_j\log\lambda_1 + m_1\log\lambda_j|
\end{align*}
where the last inequality holds using the bound $1-e^{-x}\leq x$.
In order to make this estimate independent of the choice of $x_1$, 
we define $c(v) = \min\{|v_j|:v_j \neq 0\}$ for $v\in \Rl^d$ and
bound $m_1$ and $\lambda_1^2$  below by $c(m)$ and $c(\vec{\lambda})$, respectively.

The expression $-m_j\log\lambda_1+m_1\log\lambda_j$ can be interpreted as
$\langle \log\vec\lambda,f_j\rangle$, where $f_j=-m_je_1 + m_1 e_j$,
$j=2,\ldots, d$. Let $P$ denote the orthogonal projection onto the bounding
hyperplane perpendicular to $m$.
Clearly $\langle m, f_j\rangle =0$, hence $Pf_j = f_j$.
We have $\Vert P \log \vec\lambda\Vert = \Vert  \log \vec\lambda\Vert |\sin
\theta|$.
Also, note $\Vert f_j\Vert \leq \Vert m\Vert =1$. With this, we can make the
following estimate:
$$
\sum_{j=2}^d |\langle \log\vec\lambda,f_j\rangle| = \sum_{j=2}^d |\langle  \log\vec\lambda, Pf_j\rangle| =  \sum_{j=2}^d |\langle P \log\vec\lambda, f_j\rangle|
\leq \sum_{j=2}^d \Vert P\log \vec{\lambda}\Vert \leq (d-1) \Vert \log \vec\lambda \Vert | \sin \theta |,
$$
where $\theta$ denotes the angle between $-m$ and $\log\vec\lambda$.

\begin{align*}
	\gamma(H^D) \leq \frac{2(d-1)}{c(m)c(\vec\lambda)^2}\|\log\vec{\lambda}\||\sin(\theta)|.
\end{align*}
\end{proof}

\section*{Appendix}

We now prove of Lemma \ref{lem:EpsilonCalculation}.
\begin{proof}
We bound the norm of the product of projections
$\|G^{\Lambda_{n+1}\setminus\Lambda_{n+1-\ell}}E_n\|$.  It sufficient to
consider $G^{\Lambda_{n+1}\setminus\Lambda_{n+1-\ell}}$ acting on the range of
$E_n$, that is $\caG_{\Lambda_n}\cap \caG^\perp_{\Lambda_{n+1}}$.  A vector
$\Psi$ in this subspace has the form
\begin{equation}
    \Psi = b_0\Psi_1^{\Lambda_n}\otimes\Psi_0^{\Lambda_{n+1}\setminus\Lambda_n}
    + \sum_{x\in\Lambda_{n+1}\setminus\Lambda_n}
    a_x\Psi_0^{\Lambda_n}\otimes
    \sigma_x^1\Psi_0^{\Lambda_{n+1}\backslash\Lambda_n} +
    \sum_{x\in\Lambda_{n+1}\setminus\Lambda_n}
    b_x\Psi_1^{\Lambda_n}\otimes\sigma_x^1\Psi_0^{\Lambda_{n+1}\backslash\Lambda_n} 
\end{equation}
where $b_0,\ a_x,\ b_x \in \mathbb{C}$, $\Psi_0^\Lambda,\ \Psi_1^\Lambda$ are
the zero and one particle ground states in finite set $\Lambda$, and
$\sigma_x^1\Psi_0^{\Lambda_{n+1}\backslash\Lambda_n} $ is the product state with
a single particle at site $x\in\Lambda_{n+1}\setminus\Lambda_n$.

By construction, the vector $\Psi$ is orthogonal to $\Psi_0^{\Lambda_{n+1}}$.
For $\Psi$ to be perpendicular to the single particle ground state
$\Psi_1^{\Lambda_{n+1}}$ the coefficients must satisfy the following
orthogonality condition:
\begin{equation}
    b_0 =
    \frac{-1}{\sqrt{C(\Lambda_n)}}\sum_{x\in\Lambda_{n+1}\setminus\Lambda_n}a_x\lambda^x.
\end{equation}

Since $\Lambda_{n+1}\setminus\Lambda_{n+1-\ell}$ is connected, the ground state
space is two dimensional.  The projection
$G^{\Lambda_{n+1}\setminus\Lambda_{n+1-\ell}}$ is onto the zero particle ground
state $\Psi_0^{\Lambda_{n+1}\setminus\Lambda_{n+1-\ell}}$ and the single
particle ground state, $\Psi_1^{\Lambda_{n+1}\setminus\Lambda_{n+1-\ell}}$. We
drop the notation $\Lambda_{n+1}\setminus\Lambda_{n+1-\ell}$ from the ground
state vectors when the volume is understood. The expression
$G^{\Lambda_{n+1}\setminus\Lambda_{n+1-\ell}} \Psi$ has four nonzero terms:\\
1.  The projection onto a vector with no particles in $\Lambda_{n+1-\ell}$. 
\begin{equation}
|\Psi_0\rangle\langle\Psi_0|\Psi = 
b_0\sqrt{\frac{C(\Lambda_{n+1-\ell})}{C(\Lambda_n)}}
\Psi_1^{\Lambda_{n+1-\ell}}\otimes\Psi_0^{\Lambda_{n+1}\setminus\Lambda_{n+1-\ell}}
\end{equation}
2.  The single particle projection onto the $b_0$ term.  The nonzero terms have a particle in $\Lambda_n\setminus\Lambda_{n+1-\ell}$:
\begin{equation}
|\Psi_1\rangle\langle\Psi_1|b_0\Psi_1^{\Lambda_n}\otimes\Psi_0^{\Lambda_{n+1}\setminus\Lambda_n} =
 b_0\frac{C(\Lambda_n\setminus\Lambda_{n+1-\ell})}{\sqrt{C(\Lambda_n)C(\Lambda_{n+1}\setminus\Lambda_{n+1-\ell})}}
\Psi_0^{\Lambda_{n+1-\ell}}\otimes\Psi_1^{\Lambda_{n+1}\setminus\Lambda_{n+1-\ell}}
\end{equation}
3.  The single particle projection onto the $a_x$ terms.  The nonzero terms have a particle in $\Lambda_{n+1}\setminus\Lambda_{n}$:
\begin{equation}
|\Psi_1\rangle\langle\Psi_1|\sum_{x\in\Lambda_{n+1}\setminus\Lambda_{n}}a_x|1\rangle_x =
\frac{1}{\sqrt{C(\Lambda_{n+1}\setminus\Lambda_{n+1-\ell})}}
\sum_{x\in\Lambda_{n+1}\setminus\Lambda_{n}}a_x\lambda^x
\Psi_0^{\Lambda_{n+1-\ell}}\otimes\Psi_1^{\Lambda_{n+1}\setminus\Lambda_{n+1-\ell}}
\end{equation}
4.  The single particle projection onto the $b_x$ terms.  The nonzero terms have a particle in $\Lambda_{n+1}\setminus\Lambda_{n}$ and another in $\Lambda_{n+1-\ell}$:
\begin{equation}
|\Psi_1\rangle\langle\Psi_1|\sum_{x\in\Lambda_{n+1}\setminus\Lambda_{n}}b_x\Psi_1^{\Lambda_n}\otimes|1\rangle_x =
\frac{\sqrt{C(\Lambda_{n+1-\ell})}}{\sqrt{C(\Lambda_n)C(\Lambda_{n+1}\setminus\Lambda_{n+1-\ell})}}
\sum_{x\in\Lambda_{n+1}\setminus\Lambda_{n}}b_x\lambda^x
\Psi_1^{\Lambda_{n+1-\ell}}\otimes\Psi_1^{\Lambda_{n+1}\setminus\Lambda_{n+1-\ell}}
\end{equation}
The inner product of the the sum of these vectors gives the (square) of the norm:
\begin{align*}
\|G^{\Lambda_{n+1}\setminus\Lambda_{n+1-\ell}} \Psi\|^2 & = |b_0|^2\frac{C(\Lambda_{n+1-\ell})}{C(\Lambda_n)} + \frac{C(\Lambda_{n+1-\ell})}{C(\Lambda_n)C(\Lambda_{n+1}\setminus\Lambda_{n+1-\ell})}\left|\sum_{\Lambda_{n+1}\setminus\Lambda_n}b_x\lambda^x\right|^2 \\
    &+ \left|\frac{b_0C(\Lambda_n\setminus\Lambda_{n+1-\ell})}{\sqrt{C(\Lambda_n)C(\Lambda_{n+1}\setminus\Lambda_{n+1-\ell})}} +\frac{1}{\sqrt{C(\Lambda_{n+1}\setminus\Lambda_{n+1-\ell})}}\sum_{\Lambda_{n+1}\setminus\Lambda_n}a_x\lambda^x\right|^2\\
\end{align*}
Applying the orthogonality condition for $|b_0|^2$ and combining the $a_x$ terms
simplifies 
\begin{align*}
   \frac{C(\Lambda_{n+1-\ell})}{C(\Lambda_n)^2}&\left|\sum_{\Lambda_{n+1}\setminus\Lambda_n}a_x\lambda^x\right|^2
   + \left|\frac{C(\Lambda_n) - C(\Lambda_n\setminus\Lambda_{n+1-\ell})}{C(\Lambda_n)\sqrt{C(\Lambda_{n+1}\setminus\Lambda_{n+1-\ell})}} 
	\sum_{\Lambda_{n+1}\setminus\Lambda_n} a_x \lambda^x\right|^2\\
    & = \left|\sum_{\Lambda_{n+1}\setminus\Lambda_n}a_x\lambda^x\right|^2 \left[\frac{C(\Lambda_{n+1-\ell})}{C(\Lambda_n)^2} + \frac{C(\Lambda_{n+1-\ell})^2}{C(\Lambda_n)^2C(\Lambda_{n+1}\setminus\Lambda_{n+1-\ell})} \right]\\    
    & = \left|\sum_{\Lambda_{n+1}\setminus\Lambda_n}a_x\lambda^x\right|^2 \left[\frac{C(\Lambda_{n+1-\ell})\left(C(\Lambda_{n+1}\setminus\Lambda_{n+1-\ell})+C(\Lambda_{n+1-\ell})  \right)}{C(\Lambda_n)^2C(\Lambda_{n+1}\setminus\Lambda_{n+1-\ell})} \right]\\   
        & =\left|\sum_{\Lambda_{n+1}\setminus\Lambda_n}a_x\lambda^x\right|^2 \left[\frac{C(\Lambda_{n+1-\ell})C(\Lambda_{n+1})}{C(\Lambda_n)^2C(\Lambda_{n+1}\setminus\Lambda_{n+1-\ell})} \right]\\  
        & = \left|\sum_{\Lambda_{n+1}\setminus\Lambda_n}a_x\lambda^x\right|^2 \left[\frac{C(\Lambda_{n+1-\ell})\left(C(\Lambda_{n+1}\setminus\Lambda_{n})+C(\Lambda_{n})  \right)}{C(\Lambda_n)^2C(\Lambda_{n+1}\setminus\Lambda_{n+1-\ell})} \right]\\   
    & = \left|\sum_{\Lambda_{n+1}\setminus\Lambda_n}a_x\lambda^x\right|^2 \frac{C(\Lambda_{n+1-\ell})}{C(\Lambda_n)C(\Lambda_{n+1}\setminus\Lambda_{n+1-\ell})}\left[ 1 + \frac{C(\Lambda_{n+1}\setminus\Lambda_{n})}{C(\Lambda_n)}\right]\\
\end{align*}
We combine terms and apply Cauchy-Schwarz for an upper bound:
\begin{align*}
	\|G\Psi\|^2  & = \frac{C(\Lambda_{n+1-\ell})}{C(\Lambda_n)C(\Lambda_{n+1}\setminus\Lambda_{n+1-\ell})}
    \left[\left|\sum_{\Lambda_{n+1}\setminus\Lambda_n}b_x\lambda^x\right|^2 + \left|\sum_{\Lambda_{n+1}\setminus\Lambda_n}a_x\lambda^x\right|^2 + \frac{C(\Lambda_{n+1}\setminus\Lambda_n)|\sum_{\Lambda_{n+1}\setminus\Lambda_n}a_x\lambda^x|^2}{C(\Lambda_n)}\right]\\
	& \leq \frac{C(\Lambda_{n+1-\ell})C(\Lambda_{n+1}\setminus\Lambda_n)}{C(\Lambda_n)C(\Lambda_{n+1}\setminus\Lambda_{n+1-\ell})}
    \left[\sum_{\Lambda_{n+1}\setminus\Lambda_n}|b_x|^2 + \sum_{\Lambda_{n+1}\setminus\Lambda_n}|a_x|^2 + \frac{|\sum_{\Lambda_{n+1}\setminus\Lambda_n}a_x\lambda^x|^2}{C(\Lambda_n)}\right]\\
    & = \frac{C(\Lambda_{n+1-\ell})C(\Lambda_{n+1}\setminus\Lambda_n)}{C(\Lambda_n)C(\Lambda_{n+1}\setminus\Lambda_{n+1-\ell})}\|\Psi\|^2
\end{align*}
We have equality when we choose $a_x = b_x = \lambda^x$.  Therefore,
$$
\|G^{\Lambda_{n+1}\setminus\Lambda_{n+1-\ell}}E_n\|^2
=
\frac{C(\Lambda_{n+1-\ell})C(\Lambda_{n+1}\setminus\Lambda_n)}{C(\Lambda_n) C(\Lambda_{n+1}\setminus\Lambda_{n+1-\ell})}.
$$
\end{proof}

\section*{Acknowledgements} B.N. acknowledges the stimulating environment and warm hospitality at the Erwin Schr\"odinger International Institute for Mathematical Physics, Vienna during the program {\it Quantum Many-Body Systems, Random Matrices, and Disorder}, July 2015. This research was supported in part by the National Science Foundation under Grant DMS-1515850 (A.Y. and B.N.).


\begin{thebibliography}{1}

\bibitem{bachmann:2012}
S.~Bachmann and B.~Nachtergaele.
\newblock {Product vacua with boundary states},
\newblock {\em Phys. Rev. B}, 86(3):035149, 2012.

\bibitem{bachmann:2015}
S. Bachmann, E. Hamza, B. Nachtergaele, and A. Young,
\newblock Product vacua and boundary state models in d dimensions.
\newblock {\em J. Stat. Phys.},   160: 636--658, 2015.

\bibitem{bachmann:2015a}
S. Bachmann and Y. Ogata,
\newblock $C^1$-classification of gapped parent Hamiltonians of quantum spin chains,
\newblock {\em Commun. Math. Phys.}, 338,:1011--1042, 2015.

\bibitem{cirac:2013}
J.I. Cirac, S. Michalakis, D. Perez-Garcia, and N. Schuch,
\newblock Robustness in Projected Entangled Pair States,
\newblock {\em Phys. Rev. B}, 88:115108, 2013.

\bibitem{chen:2013}
X.~Chen, Z.-C. Gu, Z.-X. Liu, and X.-G. Wen.
\newblock Symmetry protected topological orders and the group cohomology of their symmetry group,
\newblock {\em Phys. Rev. B}, 87:155114, 2013.

\bibitem{duivenvoorden:2013}
K.~Duivenvoorden and T.~Quella.
\newblock {Topological phases of spin chains}.
\newblock {\em Phys. Rev. B}, 87:125145, 2013.

\bibitem{fannes:1992}
M. Fannes, B. Nachtergaele, and R. F. Werner,
\newblock{Finitely correlated states of quantum spin chains},
\newblock {\em Commun. Math. Phys.}, 144:443--490, 1992.

\bibitem{kitaev:2003}
A.~Yu. Kitaev.
\newblock Fault-tolerant quantum computation by anyons.
\newblock {\em Ann. Phys.}, 303:2, 2003.

\bibitem{levin:2005}
M.A. Levin, and X.-G. Wen,
\newblock{String-net condensation: A physical mechanism for topological phases},
\newblock {\em Phys. Rev. B}, 71:04511, 2005

\bibitem{nachtergaele1996spectral}
B. Nachtergaele,
\newblock The spectral gap for some spin chains with discrete symmetry
  breaking.
\newblock {\em Commun. Math. Phys.}, 175(3):565--606, 1996.

\bibitem{qi:2012}
X.-L. Qi, H. Katsura, and A. W. W. Ludwig,
\newblock General Relationship Between the Entanglement Spectrum and the Edge State Spectrum of Topological Quantum States
\newblock {\em Phys. Rev. Lett.}, 108:196402, 2012.

\bibitem{perez-garcia:2010}
D. Perez-Garcia, M. Sanz, C.E. Gonzalez-Guillen, M.M. Wolf, and J.I. Cirac,
\newblock{Characterizing symmetries in a projected entangled pair state},
\newblock{\em New J. Phys.}, 12:025010, 2010.

\bibitem{Schuch:2011}
N.~Schuch, D.~Perez-Garcia, and I.~Cirac.
\newblock Classifying quantum phases using matrix product states and peps.
\newblock {\em Phys. Rev. B}, 84:165139, 2011.

\bibitem{Schuch:2013}
N.~Schuch, D.~Poilblanc, J.I. Cirac, and D.~Perez-Garcia.
\newblock Topological order in {PEPS}: Transfer operator and boundary
  {H}amiltonians.
\newblock {\em Phys. Rev. Lett.}, 111:090501, 2013.

\bibitem{verstraete:2004}
F. Verstraete, and J.I. Cirac,
\newblock{Valence-bond states for quantum computation},
\newblock {\em Phys. Rev. A}, 70:060302(R), 2004.

\bibitem{verstraete:2008}
F. Verstraete, V. Murg, and J.I. Cirac,
\newblock{Matrix product states, projected entangled pair states, and variational renormalization group methods for quantum spin systems},
\newblock {\em Advances in Physics}, 57:143--224, 2008.

\end{thebibliography}
\end{document}